\documentclass[11pt]{article}
\usepackage[a4paper, margin=20mm]{geometry}
\usepackage{amsmath, amsfonts, amssymb}
\usepackage{multirow, array}
\usepackage{xr}
\usepackage{graphicx}
\usepackage{float}
\usepackage{booktabs}
\usepackage{import}
\usepackage{hyperref}
\usepackage[numbers]{natbib}
\usepackage{pdflscape}
\usepackage{url}
\usepackage{color}
\usepackage{rotating}
\usepackage{mathabx}
\usepackage{subcaption}
\usepackage[export]{adjustbox}
\usepackage{longtable, threeparttable, makecell}
\usepackage{caption}
\usepackage{authblk}
\usepackage{setspace}

\DeclareMathAlphabet{\mathbbold}{U}{bbold}{m}{n} 

\title{Modeling Spatial Heterogeneity in Exposure Buffers and Risk: A Hierarchical Bayesian Approach} 
\author[1]{Saskia Comess}
\author[2]{Daniel E.\ Ho}
\author[3]{Joshua L.\ Warren}
\affil[1]{Emmett Interdisciplinary Program in Environment and Resources, Stanford University, Stanford, CA 94305, USA}
\affil[2]{Stanford Regulation, Evaluation and Governance Lab, Stanford Law School, Stanford University, Stanford, CA 94305, USA} 
\affil[3]{Department of Biostatistics, Yale University, New Haven, CT 06520, USA}

\date{}

\begin{document}

\maketitle

\abstract{Place-based epidemiology studies often rely on circular buffers to define ``exposure'' to spatially distributed risk factors, where the buffer radius represents a threshold beyond which exposure does not influence the outcome of interest. This approach is popular due to its simplicity and alignment with public health policies. However, buffer radii are often chosen relatively arbitrarily and assumed constant across the spatial domain. This may result in suboptimal statistical inference if these modeling choices are incorrect. To address this, we develop SVBR (Spatially-Varying Buffer Radii), a flexible hierarchical Bayesian spatial change points approach that treats buffer radii as unknown parameters and allows both radii and exposure effects to vary spatially. Through simulations, we find that SVBR improves estimation and inference for key model parameters compared to traditional methods. We also apply SVBR to study healthcare access in Madagascar, finding that proximity to healthcare facilities generally increases antenatal care usage, with clear spatial variation in this relationship. By relaxing rigid assumptions about buffer characteristics, our method offers a flexible, data-driven approach to accurately defining exposure and quantifying its impact. The newly developed methods are available in the R package \textbf{EpiBuffer}.}

\maketitle

\section{Introduction}\label{sec:background}
Epidemiological studies often seek to understand how spatially distributed features influence health outcomes \citep{Auchincloss2012}. Place-based research frequently defines an outcome unit's ``exposure" to features of interest using simple distance-based metrics, such as aggregate metrics within predefined geographic units (e.g., census blocks) or circular buffers surrounding locations of interest \citep{Auchincloss2012, Reich2021}. Common buffer-based metrics include binary presence/absence of features \citep{Towongo2022}, densities or counts \citep{Borlee2017}, and distance-weighted functions within the radius cutoff \citep{Gonzalez2020}. Uniform distance buffers are widely utilized for exposure assessment in studies of pollutant sources (e.g., animal agriculture facilities \citep{Smit2014, Borlee2017, Simoes2022}, oil and gas drilling wells \citep{Clark2022, Gonzalez2022}, pesticides \citep{Warren2014}, golf courses \citep{Krzyzanowski2025}, and other environmental hazards \citep{Chakraborty2011}), greenness \citep{Wilt2023}, and built-environment features \citep{Berke2007}. Buffers have also been used to assess patterns in neighborhood socio-economic status \citep{Patel2021}, access to healthcare facilities \citep{Towongo2022}, forest conservation planning \citep{Brouwers2010}, and for infectious disease modeling \citep{Shah2014}. Circular buffers remain a popular choice in epidemiology research, likely due to their ease of implementation, straight-forward interpretation, and comparability between studies.

When choosing a buffer radius, the researcher is selecting a spatial threshold at which exposure is thought to no longer have an effect on the outcome of interest. However, this choice is often made without rigorous prior knowledge about the spatial range of the exposure/outcome relationship \citep{Kwan2012a}. Typically the radius is chosen somewhat arbitrarily, perhaps based on the hypothesized exposure mechanism or precedent in the literature, and is assumed constant across all locations in the study \citep{Chakraborty2011}. This ``uncertain geographic context problem'' arises due to uncertainty in the true spatial extent over which an exposure influences outcomes \citep{Kwan2012a, Kwan2012b}. Previous studies have noted inconsistent buffer size choices across investigations, even when considering the same exposure-outcome relationships \citep{Chakraborty2011, Kwan2012b}. Sensitivity analyses have demonstrated that the size and shape of buffers can influence observed associations, potentially contributing to contradictory findings seen in the literature \citep{Ebisu2014, James2014}. Some studies have compared traditional circular buffers to alternative methods, such as network- and mobility-based models, and observed that methodological choices in spatial scale can bias estimated exposure-health associations \citep{Oliver2007, Wilt2023}. However, network and mobility models that rely on individual-level location and mobility information are not widely adaptable across environmental epidemiology contexts, where such data may not be available or relevant. 

Several studies have proposed statistical approaches to estimating buffer radii in place-based epidemiology studies, aiming to avoid reliance on prespecified spatial scales. One class of methods adapts distributed lag models (DLMs), traditionally used in time-series analysis, to spatial applications \citep{baek2016distributed, baek2016hierarchical, Baek2017}. In this spatial DLM framework, lags represent exposure values within concentric annuli (i.e., exposure in the region between radii $r_{l-1}$ and $r_l$) and DLM coefficients are estimated using smoothing splines. The annulus in which the coefficient trends to zero defines the maximum effect distance (i.e., the estimated buffer radius) \citep{baek2016distributed}. This method has been extended to hierarchical DLMs that allow for between-group and between-individual heterogeneity in both the spatial scale and effect magnitude \citep{baek2016hierarchical, Baek2017}. While these extensions offer greater flexibility, these models provide only indirect inference on the appropriate buffer radius via analysis of the estimated exposure effect parameters across different distances. Additionally, the need to predefine a set of discrete distance intervals in which to estimate coefficients introduces assumptions about relevant breakpoints, while replication within the individual or study region is typically required to estimate heterogeneity in the associations. 

An alternative method, developed for time-to-event studies, applies functional linear Cox regression in a two-stage process to identify the maximum radius at which an exposure effect exists \citep{Lee2024_preprint}. In the first stage, the model uses sparsity and smoothness penalties to select regions with non-zero exposure effect, and the buffer radius is taken as the supremum of the non-null regions. In the second stage, the model is refit within the identified buffer radius to estimate the functional exposure-outcome association. This approach enables a principled search for relevant distances, but like DLM-based methods, requires a discrete sequence of radii to define concentric rings. Additionally, the two-stage modeling framework means that statistical inference in the second stage does not naturally account for uncertainty introduced by selection of non-null regions in the first stage, which may impact the final inference \citep{Comess2022}.

A few studies have proposed Bayesian approaches to estimate buffer radii. \cite{Ozga2020} adapt a Bayesian change point detection model, traditionally used in time-series analysis, to a spatial setting by estimating location-specific shifts in outcome probabilities as a function of distance from an index location. While this method introduces a novel spatial application of change point analysis, it models each location independently and does not explicitly account for spatial correlation, thereby limiting information-sharing across nearby areas. Alternatively, \cite{Warren2018} apply hierarchical Bayesian methods to estimate both the buffer radius and corresponding exposure effect in a study of spillover risks of drug-resistant tuberculosis surrounding a prison. While this approach accounts for spatial correlation in the responses and allows for direct inference on the radius, it does not allow for spatial variability in any of the exposure-relevant parameters.

The primary objective of our work is to offer a data-driven alternative to the ad-hoc selection of buffer radii in epidemiology studies, while relaxing some of the surrounding decisions made in previous statistical modeling work in this area. To do so, we develop a flexible hierarchical Bayesian spatial change points model that jointly estimates the buffer radii and magnitude of the exposure effect within that distance threshold, allowing for variation in both parameters across space. Our model includes existing approaches as limiting cases, providing a unified analysis framework. Our specific contributions include, (1) we model spatial heterogeneity in buffer radii and exposure effect parameters across locations using techniques that do not require spatial replication; (2) we model the radii as continuous variables, eliminating the need for a discrete distance sequence with arbitrary cut-points; (3) we allow for direct inference on the spatially-varying radii parameters within a single model; and (4) we develop an R package (i.e., \textbf{EpiBuffer}) to facilitate future use and extensions. 

In Section \ref{sec:motivating_data} we present our motivating dataset and in Section \ref{sec:svbr_model} we detail our proposed model. We evaluate the model via a simulation study in Section \ref{sec:sim_study} and apply it to the motivating data in Section \ref{sec:application_study}. Section \ref{sec:discussion} offers conclusions and discussions of future work. 

\section{Madagascar data}\label{sec:motivating_data}
\subsection{Individual- and cluster-level data}
\begin{figure}[h]
    \centering
       \includegraphics[width=\textwidth, trim = 0 1.5cm 0 0, clip]{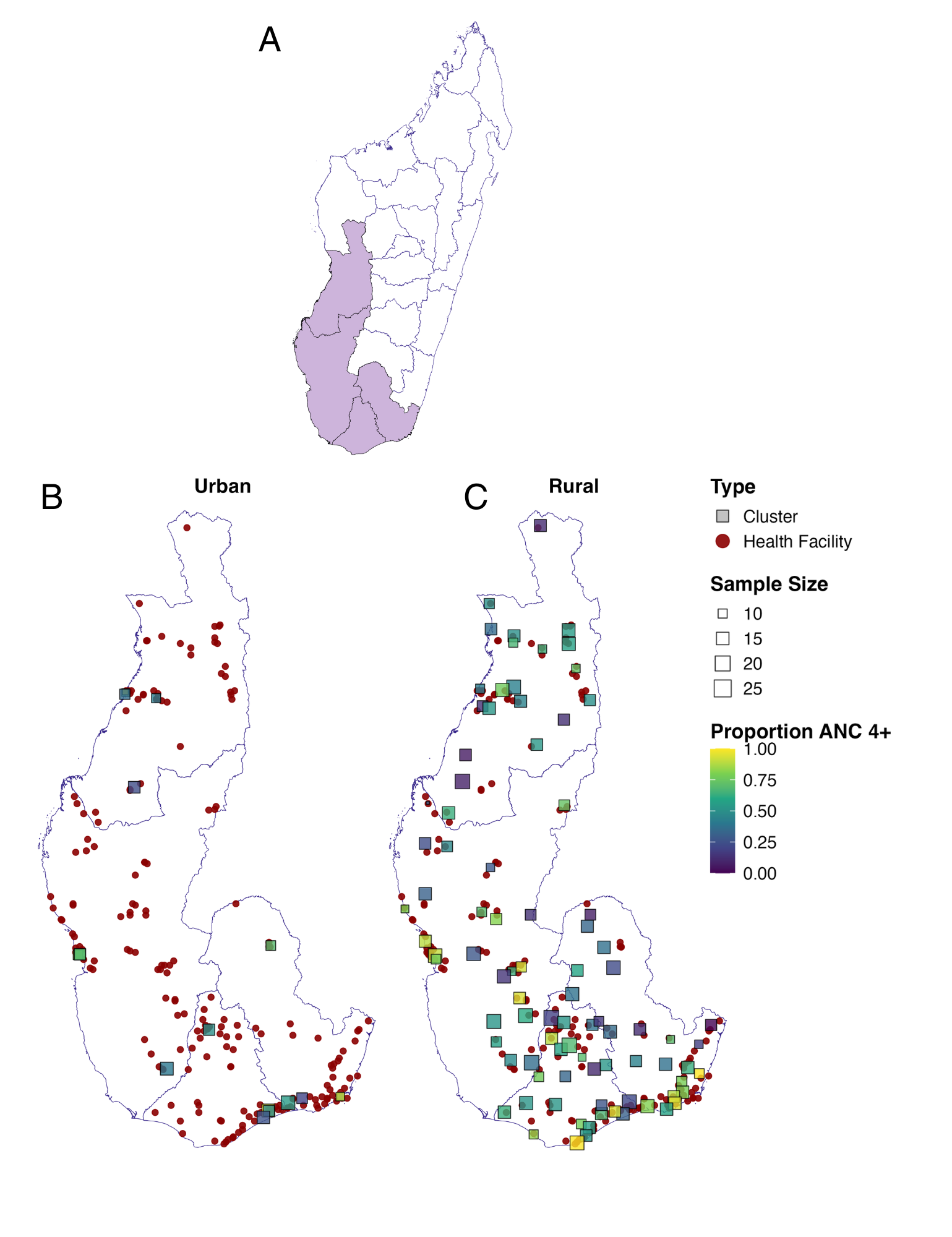}
        \caption{\small Map of Madagascar, with the study area, Toliara Province highlighted (Panel A), and map of the study area showing health facility locations (red circles) and cluster locations (squares), where cluster sample size is indicated by the size of the square and proportion of the sample that completed $\ge 4$ ANC visits is indicated by the color of the square (Panels B and C). Urban and rural designated clusters are plotted separately (Panel B, Urban; Panel C, Rural).}\label{fig:md_study_area_map}
\end{figure}

We analyze data from the 2021 Demographic and Health Surveys (DHS-8) in Madagascar, focusing on women of reproductive age (15-49 years) in the former Toliara Province (DHS regions Menabe, Atsimo-Andrefana, Androy and Anosy) (Figure \ref{fig:md_study_area_map}). The region was selected because it has historically performed poorly on maternal/child health metrics and has relatively low access to a range of critical healthcare interventions for women and children \citep{RandretsaIarivony2005, Countdown2030}. We use individual-level responses to compute demographic and health indicators relevant to pregnancy and antenatal care (ANC), using the indicators defined by DHS and the accompanying code they provide \citep{DHSv8, DHSGithub}. DHS geographic data provide latitude/longitude for each DHS cluster, where clusters consist of $25-30$ households from which individual respondents are drawn. 

From the initial sample of women aged $15-49$ living in Toliara Province ($n=3,138$), we retain those who had their most recent live-birth within the $5$ years prior to survey administration ($1,823$) and have non-missing ANC visit and distance to health facility information ($1,818$) across $109$ cluster locations. While ANC data may not be missing at random, only five women were excluded for missing data, limiting potential selection bias.

The binary dependent variable is whether or not a woman received four or more ANC visits for her most recent live-birth, which is the DHS-defined indicator of adequate healthcare utilization during pregnancy. Individual-level covariates include predisposing factors that are thought to promote or impede use of healthcare services \citep{Armah-Ansah2024}: maternal age, education, employment status, marital status, parity, and religion. Cluster-level variables capture community features affecting travel and healthcare access \citep{Armah-Ansah2024}: elevation (meters) \citep{Mayala2022}, wealth (proportion of women in the upper two quintiles of the DHS wealth index; ``richer'' and ``richest''), and the Global Human Footprint Index (GHFI, $0$: extremely rural to $100$: extremely urban), which aggregates population density, land use, infrastructure, and accessibility (including roads, railroads, and navigable rivers) \citep{Mayala2022}. Table \ref{table:sample_covariates} summarizes all variables by ANC status.

\subsection{Health facility data}
We connect the individual- and cluster-level data with a database of healthcare facilities managed by the public health sector in sub-Saharan Africa \citep{Maina2019, HealthDataSet_Snow2019}. The dataset consists of a geo-referenced inventory of public health facilities managed by governments, local authorities, faith-based organizations and non-governmental organizations. Private health facilities, those serving only special populations (e.g., prisons, schools), and those providing only specialized care are excluded. These data have previously been used to study distance to healthcare and utilization of ANC, including via linkage to DHS data \citep{Tanou2021}.

The health facility dataset contains $2,677$ facilities across Madagascar, of which $2,647$ have non-missing latitude/longitude coordinates and non-duplicate geometries. We use ArcGIS Pro and the ESRI Street Map Premium `Afrika.mmpk' product to compute the walking distances (in kilometers (kms)) on valid roads and paths between Toliara DHS clusters and healthcare facilities, allowing for a $10$ km search tolerance for snapping distance to a network feature (road or other path) {and retaining only those facilities that can be connected with a valid network feature ($n = 2,619$). Using the walking distance travel mode excludes roads and trails that are considered inaccessible to pedestrians, such as private roads or those with restrictions against walking, and allows for routes along pedestrian-only paths. Distances represent the shortest possible route between points on these valid thoroughfares \citep{EsriODMat, EsriTravelMode}.

In the resulting cluster-facility distance matrix $(109 \text{ clusters} \times 2,619 \text{ facilities})$, we remove any facilities with a missing distance to one or more clusters. Missing values may occur if there is not a valid route between locations or if the point cannot be located on the map. This results in a final distance matrix containing $109$ clusters and $2,598$ health facilities. 

\subsection{Survey design}
DHS data are collected using a two-stage survey sample design with varying selection probabilities across respondents \citep{DHSv8}. Incorporating survey weights into hierarchical Bayesian models remains an active area of methodological research \citep{Si2015, Si2020, Cassy2022}. In the absence of a well-established weighting procedure for hierarchical Bayesian models, a common recommendation is to include key variables related to sampling and non-response directly in the regression model \citep{Gelman2007}. For the DHS data, the primary first-stage stratifying variables are cluster-level urban/rural designation and DHS region \citep{DHSv8}. Accordingly, we include these variables in our models. Urban/rural designation and regional boundaries are both drawn from Madagascar's own administrative definitions, where regions are based on census levels and urban areas are those with $5,000$ or more inhabitants \citep{Fish2020, DHSv8}. In Section \ref{sec:sens}, we present a sensitivity analysis comparing our covariate-adjustment approach with a frequentist survey-weighted analysis, assessing robustness with respect to this alternative treatment of the survey design.

\section{Methods}\label{sec:svbr_model}
We develop the Spatially-Varying Buffer Radii (SVBR) model to estimate location-specific radii and exposure effect parameters, allowing for increased flexibility over existing approaches for defining and quantifying the impact of exposure on an outcome. Spatial variability in the parameters is modeled via covariates and random effects to accommodate known and unknown sources of heterogeneity, respectively. 

To begin, the total study population consists of $\sum_{j=1}^m n_j$ outcome units (e.g., individuals) distributed across $m$ unique spatial locations, $\textbf{s}_j$, for $j = 1,...,m$, where $n_j \geq 1$ is the number of units at location $\textbf{s}_j$. The dependent variable (e.g., health outcome) measured for unit $i$ at location $\textbf{s}_j$ is denoted by $Y_i\left(\textbf{s}_j\right)$ and is modeled as a function of exposure surrounding location $\textbf{s}_j$ and individual- and location-specific covariates. We introduce SVBR generally here with respect to outcome type to emphasize the fact that we have developed the methodology and software for several different distributions, including Gaussian, binomial, and negative binomial responses. Therefore, SVBR is given as
    \begin{equation}\label{eq:svbr}
        \begin{array}{c}
            Y_i\left(\textbf{s}_j\right) | \mu_i\left(\textbf{s}_j\right), \boldsymbol{\zeta} \stackrel{\text{ind}}{\sim} f\left(y \mid \mu_i\left(\textbf{s}_j\right), \boldsymbol{\zeta} \right),\ j=1,\hdots,m, \ i=1,\hdots,n_j, \  \\[1.5ex]
            
            g\left\{\mu_i\left(\textbf{s}_j\right)\right\} = \text{O}_i\left(\textbf{s}_j\right) + \textbf{x}_i\left(\textbf{s}_j\right)^{\text{T}} \boldsymbol{\beta} + \text{z}\left\{\textbf{s}_j; \delta\left(\textbf{s}_j\right)\right\} \theta(\textbf{s}_j), \\[1.5ex] 
        \end{array}
    \end{equation}
where $f(.|.)$ is the selected probability density function of the outcome with mean $\mu_i\left(\textbf{s}_j\right)$ and additional likelihood-defining parameters $\boldsymbol{\zeta}$; $\text{O}_i\left(\textbf{s}_j\right)$ is an optional offset term (often used when modeling count data, but zero otherwise); and $\textbf{x}_i\left(\textbf{s}_j\right)$ is a $p_x$ length vector of outcome unit- and location-specific covariates, including an intercept term, with corresponding regression coefficients $\boldsymbol{\beta}$. The exposure (i.e., $\text{z}\left\{\textbf{s}_j; \delta\left(\textbf{s}_j\right)\right\}$), buffer radius (i.e., $\delta\left(\textbf{s}_j\right)$), and exposure effect parameter (i.e., $\theta(\textbf{s}_j)$) are all location-specific and have additional definitions/models to allow for increased flexibility while maintaining parameter identifiability. 

\subsection{Exposure definitions}\label{sec:exposure_definitions}
The exposure value at location $\textbf{s}_j$ is denoted by $\text{z}\left\{\textbf{s}_j; \delta\left(\textbf{s}_j\right)\right\}$ and is defined as a function of the location, the location(s) of the exposure source(s) (i.e., $\textbf{c}_k$, for sources $k = 1,\hdots, h$), and the location-specific buffer radius parameter, $\delta\left(\textbf{s}_j\right)$. The distance between outcome unit location $\textbf{s}_j$ and source location $\textbf{c}_k$ is given as $\text{d}_{jk} = d(\boldsymbol{s}_j, \textbf{c}_k)$, where $d(.,.)$ can be any appropriate measure of distance (e.g., Euclidean, Manhattan, road-network). We consider two representative exposure definitions:
\begin{itemize}
\begin{subequations}
    \item \textbf{Counts} of sources within $\delta\left(\textbf{s}_j\right)$ of $\textbf{s}_j$:
    \begin{equation}\label{eq:exposure_counts}
        \text{z}\left\{\textbf{s}_j; \delta\left(\textbf{s}_j\right)\right\} = \sum_{k=1}^h \mathbbold{1}\{\text{d}_{jk} \leq \delta\left(\textbf{s}_j\right)\}; \text{ and }
    \end{equation}
    \item \textbf{Spherical exposure}, where the range and weighting of exposure sources is defined completely by $\delta\left(\textbf{s}_j\right)$:
    \small{
    \begin{equation}\label{eq:exposure_spherical}
        \text{z}\left\{\textbf{s}_j; \delta\left(\textbf{s}_j\right)\right\} =
        \sum\limits_{k=1}^{h} \mathbbold{1}\{\text{d}_{jk} \leq \delta\left(\textbf{s}_j\right)\} \left[1 - 1.5\left\{\frac{\text{d}_{jk}}{\delta\left(\textbf{s}_j\right)}\right\} + 0.5\left\{\frac{\text{d}_{jk}}{\delta\left(\textbf{s}_j\right)}\right\}^3 \right],
    \end{equation}
    }
\end{subequations}
\end{itemize} 
where $\mathbbold{1}\{.\}$ denotes an indicator function equal to one if the input statement is true and equal to zero otherwise. The spherical definition assigns continuous weights in $\left[0,1\right]$ to each source location, with those closer to $\textbf{s}_j$ receiving higher weights. Any source locations further than $\delta\left(\boldsymbol{s}_j\right)$ away automatically receive a weight of zero. This weighting function resembles the commonly used inverse distance weighting (IDW) approach in that sources closer to an outcome location contribute more to the total exposure than those further away. However, it is more flexible and interpretable because $\delta\left(\textbf{s}_j\right)$ explicitly controls both the rate of decay of the weights \textbf{and} the maximum distance at which a source contributes to the total exposure (i.e., beyond $\delta\left(\textbf{s}_j\right)$ the weight is zero). In contrast, IDW uses a fixed power parameter that does not directly specify a cutoff distance for the local range of influence.  

Both exposure definitions use $\delta\left(\textbf{s}_j\right)$ to effectively divide the outcome locations into two groups; those with zero exposure (i.e., no source locations within the buffer) and those with non-zero exposure. This division is important, as the unexposed observations establish the baseline behavior, while the exposed observations quantify the departure from baseline, with both needed to estimate $\theta(\textbf{s}_j)$.

\subsection{Spatially-varying radii and exposure effect parameters}\label{sec:svbr_radii_theta_specifications}
We introduce a model for the spatially-varying buffer radii, $\delta\left(\textbf{s}_j\right)$, that is a function of location-specific predictors $\textbf{w}(\textbf{s}_j)$ (with $p_w$ predictors, including an intercept) and spatially correlated random effects $\phi(\textbf{s}_j)$. Because $\delta\left(\textbf{s}_j\right)$ are bounded by fixed $a$ (typically, $a=0$) and $b>0$, we apply a transformation before introducing the regression model.  Specifically, we use the inverse cumulative distribution function (CDF) of the standard normal distribution (i.e., $\Phi^{-1}\left(.\right)$) such that  
    \begin{subequations}
    \begin{equation}\label{eq:spatial_radii}
        \Phi^{-1} \left( \frac{\delta\left(\textbf{s}_j\right) - a}{b - a}\right) = \textbf{w}(\textbf{s}_j)^{\text{T}} \boldsymbol{\gamma} + \phi(\textbf{s}_j)
    \end{equation} where $\boldsymbol{\gamma}$ is the vector of regression parameters, and the $\phi(\textbf{s}_j)$ parameters are modeled jointly using a Gaussian process prior distribution centered at a vector of zeros with spatially referenced exponential covariance structure based on the Euclidean distance between locations such that $\boldsymbol{\phi} = \left\{\phi\left(\textbf{s}_1\right), \hdots, \phi\left(\textbf{s}_m\right)\right\}^{\text{T}} | \tau_{\phi}^2, \rho_{\phi} \sim \text{MVN}\left(\textbf{0}_m, \tau_{\phi}^2\Sigma(\rho_{\phi})\right)$ and $\Sigma\left(\rho_{\phi}\right)_{jj'} = \exp\left\{-\rho_{\phi} || \textbf{s}_{j} - \textbf{s}_{j'}||\right\}$ \citep{banerjee2003hierarchical}. The marginal variance of the spatial process is denoted by $\tau^2_{\phi}$; the spatial correlation parameter is denoted by $\rho_{\phi}$ and describes the rate of decay in correlation as a function of distance (i.e., larger $\rho_{\phi}$ implies correlation between points decreases more quickly as distance increases); $||.||$ represents the Euclidean distance function; and $\textbf{0}_m$ is an $m$-length vector of zeros.
    
The location-specific parameter describing the exposure effect (i.e., $\theta(\textbf{s}_j)$) is modeled as a function of $p_q$ (including an intercept) predictors contained in the $\textbf{q}(\textbf{s}_j)$ vector. Because we anticipate the same predictors could impact the buffer radii and the exposure effects, we recommend defining $\textbf{q}(\textbf{s}_j) \equiv \textbf{w}(\textbf{s}_j)$, although this is not a formal requirement of the model. Specifying the exposure effect as a low-dimensional function of location-specific predictors helps maintain identifiability in the setting where having both the buffer radii and exposure effect as unknown parameters may result in excessive flexibility. Specifically, we define $\theta(\textbf{s}_j)$ as 
\begin{equation}\label{eq:theta_polynomial}
    \theta(\textbf{s}_j) = \textbf{q}(\textbf{s}_j)^{\text{T}} \boldsymbol{\eta},
\end{equation} where $\boldsymbol{\eta}$ are the corresponding regression parameters. This is equivalent to including interactions between the exposure and predictors from $\textbf{q}\left(\textbf{s}_j\right)$ in the primary regression model from (\ref{eq:svbr}).
\end{subequations}

\subsection{Nested SVBR models}\label{sec:svbr_special_cases}
Within the general SVBR framework, there are several limiting cases that are of particular interest. For the purposes of comparing the different versions of the model, we will refer to the full SVBR framework as SVBR$\left[\delta\left(\textbf{s}_j\right), \theta(\textbf{s}_j)\right]$ to emphasize that both sets of parameters are varying spatially.

One of the simplest variants is a model that includes only one buffer radius parameter and one exposure effect parameter (i.e., SVBR$\left[\delta, \theta\right]$). SVBR$\left[\delta, \theta\right]$ is a natural extension of the models typically fit in the epidemiology literature that assume a single radius; however, in SVBR$\left[\delta, \theta\right]$, the buffer radius is treated as an unknown parameter instead of being fixed \textit{a priori}. SVBR$\left[\delta, \theta\right]$ is obtained by omitting the spatial random effects (i.e., $\tau^2_{\phi} \equiv 0$ such that $\phi(\textbf{s}_j) \equiv 0$ for all $j$) from (\ref{eq:spatial_radii}) and defining the location-specific covariates as $\text{w}(\textbf{s}_j) = \text{q}\left(\textbf{s}_j\right) = 1$ for all $j$ with $p_w = p_q = 1$.  As a result, we obtain $\delta = \left(b-a\right)\Phi(\gamma_0) + a$ and $\theta = \eta_0$, where both parameters no longer depend on a specific spatial location.  Further, when $\gamma_0$ is given a $\text{N}\left(0,1\right)$ prior distribution, from the probability integral transform we know that $\Phi\left(\gamma_0\right) \sim \text{Uniform}\left(0,1\right)$ and therefore, $\delta \sim \text{Uniform}\left(a,b\right)$; an intuitive prior distribution for the buffer radius.

When $\textbf{q}(\textbf{s}_j) = 1$ for all $\textbf{s}_j$ and $p_q=1$ (i.e., intercept only)} in (\ref{eq:theta_polynomial}), the full model collapses to SVBR$\left[\delta\left(\textbf{s}_j\right), \theta\right]$, a version that allows for location-specific radii but a spatially-constant exposure effect since $\theta = \eta_0$.  SVBR$\left[\delta\left(\textbf{s}_j\right), \theta\right]$ is appropriate when one assumes the effect of exposure on the outcome is constant, while the effective range of exposure may differ spatially. Table \ref{table:model_desc} provides a summary of the notation for the different versions of SVBR.

\subsection{Prior specification and parameter identifiability}\label{sec:svbr_priors}
To fully specify the model, we assign weakly informative prior distributions to the non-radii regression parameters such as $\beta_j \stackrel{\text{iid}}{\sim} \text{N}\left(0, 100^2\right)$, $j=0,\hdots,p_x-1$ and $\eta_j \stackrel{\text{iid}}{\sim} \text{N}\left(0, 100^2\right)$, $j=0,\hdots,p_q-1$. The prior distribution for the spatial correlation parameter is given as $\rho_{\phi} \sim \text{Gamma}(1, 1)$, and is selected based on the fact that we scale the spatial distances to range between $0$ and $1$ prior to analysis. 

To further improve identifiability of the spatially-varying radii, we specify prior distributions for $\gamma_j$ that encourage a clear partitioning of variability between the predictor-driven and spatial components. Specifically, the priors are given as $$\gamma_j | \tau^2_{\phi} \stackrel{\text{iid}}{\sim} \text{N}\left(0, \frac{1-\tau^2_{\phi}}{p_w}\right),\ j=0,\hdots,p_w-1,$$ with $\tau_{\phi} \sim \text{Uniform}\left(0,1\right)$ the marginal standard deviation of the spatially-correlated random effects from (\ref{eq:spatial_radii}). We note that for SVBR$\left[\delta, \theta\right]$ where $p_w =1$ and $\tau_{\phi} = 0$, this results in $\gamma_0 \sim \text{N}\left(0,1\right)$ (as described in Section \ref{sec:svbr_special_cases}). 

Given this form, the prior conditional variance of the transformed $\delta\left(\textbf{s}_j\right)$ parameter shown in (\ref{eq:spatial_radii}) is $$\text{Var}\left(\textbf{w}\left(\textbf{s}_j\right)^{\text{T}}\boldsymbol{\gamma} + \phi\left(\textbf{s}_j\right) | \textbf{w}\left(\textbf{s}_j\right), \tau^2_{\phi}\right) = \frac{1-\tau^2_{\phi}}{p_w} \sum_{k=0}^{p_w-1} \text{w}_k\left(\textbf{s}_j\right)^2 + \tau^2_{\phi}.$$ With predictors in $\textbf{w}\left(\textbf{s}_j\right)$ standardized to have a mean of zero and variance of one prior to analysis (and the intercept equal to one), averaging over $\textbf{w}\left(\textbf{s}_j\right)$ gives \begin{align*}\text{E}_{\textbf{w}}\left\{\text{Var}\left(\textbf{w}\left(\textbf{s}_j\right)^{\text{T}}\boldsymbol{\gamma} + \phi\left(\textbf{s}_j\right) | \textbf{w}\left(\textbf{s}_j\right), \tau^2_{\phi}\right)\right\}  & = \frac{1 - \tau^2_{\phi}}{p_w}\sum_{k=0}^{p_w-1} \text{E}_{\text{w}}\left\{\text{w}^2_k\left(\textbf{s}_j\right)\right\} + \tau^2_{\phi}\\ & \approx 1 - \tau^2_{\phi} + \tau^2_{\phi} = 1.\end{align*} Because of the $\Phi^{-1}\left(.\right)$ transformation and zero-meaned Gaussian distributions used in \eqref{eq:spatial_radii}, this form implies an approximate $\text{Uniform}\left(a,b\right)$ marginal prior distribution for each of the buffer radii, which helps to avoid overfitting. Additionally, sharing the single variance parameter $\tau^2_{\phi}$ between the fixed and random effects forces the model to partition total variability between the two components and improves identifiability of the radii, since both terms cannot inflate simultaneously.

Details of the Markov chain Monte Carlo (MCMC) algorithm, full conditional distributions, and a large‑domain spatial approximation appear in Section S1 of the online supplementary materials. While the approximation was not used in this work, it is included in our \textbf{EpiBuffer} R package for those working with a larger number of unique spatial locations (i.e., larger $m$).

\section{Simulation study}\label{sec:sim_study}
We design a simulation study to explore the properties of SVBR and several nested modeling approaches across different data generating settings. The specific objectives are to evaluate how well the models infer key parameters and to assess whether several common Bayesian model comparison tools, including Watanabe-Akaike information criteria (WAIC) \citep{Watanabe2010}, Pareto-smoothed importance sampling leave-one-out (PSIS-LOO) cross validation \citep{vehtari2017practical}, and the logarithm of the pseudo marginal likelihood (LPML) \citep{geisser1979predictive}, can consistently correctly identify the most appropriate model based on the true underlying data generating process. Simulated data are based on our Madagascar case study data to ensure that the findings are relevant to our application. 

\subsection{Study design}
We motivate our simulation study using the actual healthcare facility locations and DHS data from the Toliara Province (described in Section \ref{sec:motivating_data}). For a single simulated dataset, we first randomly sample $500$ women across $m^*$ unique cluster locations from the DHS data and use their observed information for cluster location (including GHFI value: $\text{w}_1\left(\textbf{s}_j\right)$, and elevation: $\text{w}_2\left(\textbf{s}_j\right)$), distance to healthcare facilities, and the employment status covariate (i.e., $\text{x}_i\left(\textbf{s}_j\right)$). We denote the number of unique locations as $m^*$ (instead of $m$) to emphasize that this value potentially changes across simulated datasets depending on the sampled individuals. Using the randomly selected women and their corresponding data, we simulate a binary outcome for each individual from a Bernoulli distribution with logit link function that connects distance to healthcare facility with the probability of the outcome occurring. Specifically, based on (\ref{eq:svbr}) we define  
\begin{align*}
    &Y_i(\textbf{s}_j) | p_i(\textbf{s}_j) \stackrel{\text{ind}}{\sim} \text{Bernoulli}\left\{p_i(\textbf{s}_j)\right\};\ j=1,\hdots,m^*,\ i=1,\hdots,n_j,\ \sum_{j=1}^{m^*} n_j = 500;\\ &\text{logit}\left\{p_i\left(\textbf{s}_j\right)\right\} = \beta_0 + \beta_1 \text{x}_i\left(\textbf{s}_j\right) + \text{z}\left\{\textbf{s}_j; \delta\left(\textbf{s}_j\right)\right\} \theta(\textbf{s}_j),
\end{align*} 
where $\text{z}\left\{\textbf{s}_j; \delta\left(\textbf{s}_j\right)\right\}$ is the number of healthcare facilities within $\delta\left(\textbf{s}_j\right)$ km of location $\textbf{s}_j$ (i.e., exposure definition (\ref{eq:exposure_counts})) and the definitions/specifications of $\delta\left(\textbf{s}_j\right)$ and $\theta(\textbf{s}_j)$ vary by simulation setting. 

We consider five distinct simulation settings with increasing flexibility in the exposure/response relationship: 
\begin{enumerate}
    \item No Effect:  $$\theta(\textbf{s}_j) = 0 \text{ for all } j;$$
    
    \item Single Radius \& Effect:  $$\delta\left(\textbf{s}_j\right) = \left(b - a\right) \Phi\left(\gamma_0\right) + a \text{ and } \theta(\textbf{s}_j) = \eta_0 \text{ for all } j;$$
    
    \item Varying Radii, Single Effect:  $$\delta\left(\textbf{s}_j\right) = \left(b - a\right) \Phi\left\{\gamma_0 + \gamma_1 \text{w}_1\left(\textbf{s}_j\right) + \phi(\textbf{s}_j)\right\} + a;\ \theta(\textbf{s}_j) = \eta_0 \text{ for all } j;$$
    
    \item Varying Radii \& Effect:  \begin{align*}&\delta\left(\textbf{s}_j\right) = \left(b - a\right) \Phi\left\{\gamma_0 + \gamma_1 \text{w}_1\left(\textbf{s}_j\right) + \phi(\textbf{s}_j)\right\} + a;\\ 
    &\theta(\textbf{s}_j) = \eta_0 + \eta_1 \text{w}_1(\textbf{s}_j);\end{align*}
    
    \item Misspecified:
    \begin{align*}&\delta\left(\textbf{s}_j\right) = \frac{b-a}{1 + \exp\left[-\left\{\gamma_0 + \gamma_1 \text{w}_1\left(\textbf{s}_j\right) + \gamma_2 \text{w}_2\left(\textbf{s}_j\right) + \zeta\left(\textbf{s}_j\right)\right\}\right]};\\ &\theta(\textbf{s}_j) = \pi_0 + \pi_1\text{logit}\left\{\frac{\delta\left(\textbf{s}_j\right)}{b-a}\right\}.\end{align*}
\end{enumerate} Settings 2-4 correspond to SVBR$\left[\delta, \theta\right]$, SVBR$\left[\delta\left(\textbf{s}_j\right), \theta\right]$, and SVBR$\left[\delta\left(\textbf{s}_j\right), \theta(\textbf{s}_j)\right]$, respectively. Setting 5 represents a misspecified data generating algorithm with respect to the proposed model(s). Specifically, an inverse-logit link function is used to connect predictors to the $\delta\left(\textbf{s}_j\right)$ parameters instead of $\Phi\left(.\right)$; an additional covariate (i.e., $\text{w}_2\left(\textbf{s}_j\right)$) is used for data generation but is not included during model fitting (i.e., unmeasured confounding); the spatially correlated $\zeta\left(\textbf{s}_j\right)$ parameters are generated using thin-plate splines with $30$ degrees of freedom based on the latitude/longitude of each selected cluster location \citep{wood2003thin, wood2017generalized} instead of a spatially structured Gaussian process; and $\theta(\textbf{s}_j)$ is modeled as a linear function of the corresponding $\delta\left(\textbf{s}_j\right)$ parameter itself instead of the fixed predictors. This setting tests the ability of our framework to adapt when underlying assumptions have been violated.

Prior to analysis, we hypothesized that all models should perform equally well in Setting 1 where there is no association between healthcare access and the outcome. SVBR$\left[\delta, \theta\right]$ should be most efficient in Setting 2 given that it is correctly specified, though the others should be flexible enough to account for the lack of variation in the parameters. In Setting 3, SVBR$\left[\delta, \theta\right]$ should struggle since the constant radius assumption is violated while the other models should perform similarly. In Setting 4, only SVBR$\left[\delta\left(\textbf{s}_j\right), \theta(\textbf{s}_j)\right]$ should be flexible enough to capture heterogeneity in both radii and effect size parameters. Given the level of misspecification in Setting 5, the estimates for some of the parameters may be biased, but we anticipate that SVBR$\left[\delta\left(\textbf{s}_j\right), \theta(\textbf{s}_j)\right]$ may have the flexibility to estimate the radii and exposure effect parameters robustly, particularly in terms of the combined exposure effect (i.e., $\text{z}\left\{\textbf{s}_j; \delta\left(\textbf{s}_j\right)\right\}  \theta(\textbf{s}_j)$).

To further ensure realistic data are generated, the true model parameter values in each setting are selected based on the posterior distributions from fitting all forms of SVBR to the real data, where the radius is bounded by $\left(a = 0\text{ km}, b = 20 \text{ km}\right)$ (see Section \ref{sec:application_study} for application study details). Table S1 (online supplementary materials) describes the true parameter values and how they were obtained. In each setting, we simulate $500$ datasets for analysis. 

 In order to assess the impact of the likelihood choice on the final results, we repeat the entire simulation study as decribed while generating Gaussian instead of Bernoulli outcomes. Specifically, we simulate the data as $Y_i\left(\textbf{s}_j\right) \stackrel{\text{ind}}{\sim}\text{N}\left(\beta_0 + \beta_1 \text{x}_i\left(\textbf{s}_j\right) + \text{z}\left\{\textbf{s}_j; \delta\left(\textbf{s}_j\right)\right\} \theta(\textbf{s}_j), 1\right)$ from each of the settings.

\subsection{Evaluating model performance}
We apply each model (i.e., SVBR$\left[\delta, \theta\right]$, SVBR$\left[\delta\left(\textbf{s}_j\right), \theta\right]$, and SVBR$\left[\delta\left(\textbf{s}_j\right), \theta(\textbf{s}_j)\right]$) to every simulated dataset and collect information needed to evaluate their ability to estimate key model parameters, including the covariate regression parameters, radii, exposure effect parameters, and total exposure impact, which is defined as the product of the buffer-defined exposure and the exposure effect (i.e., $\text{z}\left\{\textbf{s}_j; \delta\left(\textbf{s}_j\right)\right\}  \theta(\textbf{s}_j)$). This final quantity is important to monitor to determine how the radii and exposure effect estimates behave jointly, which is particularly useful in Setting 1 where there is no impact of exposure on the outcome.  

For each target, we compute bias and mean absolute error (MAE) of the posterior median estimators along with the empirical coverage (EC) of the $95$\% highest posterior density intervals (HDI) and the length of these intervals. For targets with multiple true values, we compute the average bias, MAE, EC, and HDI length. When a model only provides a single estimate and HDI for a parameter which is actually varying across space (e.g., SVBR$\left[\delta, \theta\right]$ in Settings 3-5), we still compute the average comparison metrics while using the same single estimate/HDI for each varying parameter.

We also compare the competing methods in each setting by computing WAIC, PSIS-LOO, and LPML. Each metric provides an estimate of out-of-sample predictive accuracy based on the model's posterior distribution. As is standard, we transform each measure so that smaller values indicate an improved model fit. We are particularly interested in seeing if these tools can correctly identify the correct model that corresponds to each of the different simulation settings. If so, this could guide future applications of the methodology and help users understand which set of results to base their final inference on in practice. 

\subsection{Results}
For each dataset, we run our MCMC algorithms for $75,000$ iterations, discard the first $35,000$ as burn-in period prior to model convergence, and thin the remaining samples by a factor of $4$ to reduce posterior auto-corelation. This results in a total of $10,000$ posterior samples collected for each model fit. We monitor the Metropolis acceptance rates for all applicable parameters and the posterior standard deviation of the $\eta_j$ parameters. In rare instances (fewer then $5$ per $500$ datasets), we observed poor mixing for some of the $\eta_j$ parameters, as evidenced by exceptionally large posterior standard deviations. In these cases, adjusting the Gaussian prior distributions for $\eta_j$ to have a variance of one resolved the convergence issues. 

The MAE results are shown in Table \ref{table:sim_mse}, with results for bias, EC, and HDI length presented in the online supplementary materials (Tables S2-S4, respectively). Overall, results confirm our hypotheses regarding model performance under different levels of flexibility. The correctly specified model performs best in each setting, but overparameterized models (i.e., those with greater flexibility than required by the true data generating process) maintain good performance and can adapt to simpler structures. In contrast, underparameterized models that are too restrictive show substantial deficiencies when the true structure includes spatial variation.

In Setting 1 (No Effect), all models perform similarly on MAE and EC for the total exposure impact, correctly recognizing the null exposure association. This demonstrates that SVBR's flexibility does not lead to spurious findings when no effect exists. In Setting 2 (Single Radius \& Effect), SVBR$\left[\delta, \theta\right]$ performs best as expected for the correctly specified model. However, the more flexible models also show adequate, though less efficient, performance overall.

In Setting 3 (Varying Radii, Single Effect), SVBR$\left[\delta, \theta\right]$ clearly underperforms with larger bias and MAE, and poor EC across multiple parameters. Both SVBR$\left[\delta\left(\textbf{s}_j\right), \theta\right]$ and SVBR$\left[\delta\left(\textbf{s}_j\right), \theta\left(\textbf{s}_j\right)\right]$ capture the spatial variation in radii effectively, with similar performance metrics. In Setting 4 (Varying Radii \& Effect), only SVBR$\left[\delta\left(\textbf{s}_j\right), \theta\left(\textbf{s}_j\right)\right]$ is flexible enough to fully capture heterogeneity in both the radii and exposure effects. 

In Setting 5 (Misspecified), despite substantial model misspecification, SVBR$\left[\delta\left(\textbf{s}_j\right), \theta\right]$ and SVBR$\left[\delta\left(\textbf{s}_j\right), \theta\left(\textbf{s}_j\right)\right]$ demonstrate robustness. Both perform similarly in terms of bias and MAE, with SVBR$\left[\delta\left(\textbf{s}_j\right), \theta\left(\textbf{s}_j\right)\right]$ having slightly improved EC overall. This is likely due to the fact that the data were simulated with variability in both radii and exposure effects based on our application data results, where SVBR$\left[\delta\left(\textbf{s}_j\right), \theta\right]$ was the preferred model. These results suggest that the flexible framework can adapt when underlying assumptions are violated.

Model comparison metrics (Table S5) strongly align with these findings. In Setting 1, no model is clearly favored. SVBR$\left[\delta, \theta\right]$ dominates in Setting 2 ($66\%$ of datasets). In Settings 3-4, models allowing for spatial variation in radii are strongly preferred ($94-98\%$ of datasets), with SVBR$\left[\delta\left(\textbf{s}_j\right), \theta\left(\textbf{s}_j\right)\right]$ favored in 52--55\% of Setting 4 datasets. In Setting 5, SVBR$\left[\delta\left(\textbf{s}_j\right), \theta\right]$ and SVBR$\left[\delta\left(\textbf{s}_j\right), \theta\left(\textbf{s}_j\right)\right]$ perform similarly ($50\%$ and $32-33\%$, respectively). WAIC, PSIS-LOO, and LPML show nearly identical agreement across all settings. Overall, these metrics reliably identify the model that best describes the underlying data-generating process.

Results from the Gaussian likelihood version of the simulation study (Tables S6--S10) closely mirror the Bernoulli findings in terms of the overall patterns, but show improved performance across all metrics. Coverage probabilities are closer to the nominal 95\% level, bias is reduced, and parameter estimates are more precise. Importantly, the key patterns (i.e., improved performance of flexible models when spatial variation exists, and reliable model selection via WAIC, PSIS-LOO, and LPML) hold consistently across both likelihood choices, demonstrating that SVBR's fundamental behavior is robust to outcome type.

\section{Antenatal care and distance to healthcare facilities in Madagascar}\label{sec:application_study}
Receiving adequate healthcare during pregnancy is critical for maternal and newborn health. World Health Organization (WHO) guidelines recommend a minimum of four ANC visits during pregnancy, although recent guidance advises that completing eight or more visits provides greater benefit \citep{WorldHealthOrganization2016}. A number of studies have investigated the benefits of completing four or more ANC visits for women in low- and middle-income countries (LMICs). In Bangladesh, pregnant women who received adequate ANC were $79\%$ less likely to experience adverse perinatal outcomes, such as preterm birth, fetal distress, low birth weight, intrauterine growth retardation, admission to the neonatal care unit, and perinatal death, as compared to those who received inadequate ANC \citep{Mina2023}. Thus, improving healthcare access during pregnancy is a key global health priority and there is significant interest in understanding factors that facilitate greater ANC usage \citep{WorldHealthOrganization2016}. 

Several studies have found that access to healthcare, operationalized in terms of distance between maternal residence and healthcare facilities, is an important factor in determining frequency, timing, and quality of ANC and other pregnancy-related health indicators in LMICs \citep{Tanou2019, Tanou2021, Towongo2022, Merid2024}. However, these studies have not used methods that allow for direct inference on the spatial extent of healthcare exposure. Instead they have relied on standard regression models with distance-based predictors including continuous distance \citep{Tanou2019, Tanou2021}, binary distance thresholds such as $\leq 5$ km \citep{Ashiagbor2020, Tanou2021, Towongo2022}, and self-reported perception of distance as a barrier to receiving care \citep{Merid2024}. 

We apply SVBR to allow for a more nuanced understanding of the potentially spatially-varying effects of health facilities on ANC utilization in the Toliara Province of Madagascar (see Section \ref{sec:motivating_data} for description of the data). Our primary outcome of interest is whether a woman completed at least four ANC visits during pregnancy (binary outcome: $\ge 4$ visits (1) vs. $\textless 4$ visits (0)). We model the probability of completing $\ge 4$ ANC visits as a function of walking distance to health facilities on valid roads/paths from the woman's residence cluster; individual-level socio-demographic covariates including maternal age, marital status, education, employment, parity, and religion; and cluster-level covariates that account for the DHS survey design (i.e., urban/rural designation and DHS region). 

When modeling the radius in (\ref{eq:spatial_radii}) and exposure effects in (\ref{eq:theta_polynomial}), we include elevation, wealth, and GHFI as cluster-specific predictors (i.e., $\textbf{w}$ and $\textbf{q}$). We investigate the definitions of health facility ``exposure'' described in (\ref{eq:exposure_counts}) and (\ref{eq:exposure_spherical}) by fitting separate models where $\text{z}\{\textbf{s}_j; \delta\left(\textbf{s}_j\right)\}$ is specified in terms of facility counts and the spherical definition, respectively.

We fit SVBR$\left[\delta, \theta\right]$, SVBR$\left[\delta\left(\textbf{s}_j\right), \theta\right]$, and SVBR$\left[\delta\left(\textbf{s}_j\right), \theta(\textbf{s}_j)\right]$ on the full sample of $\sum_{j=1}^m n_j = 1,818$ individuals from $m=109$ clusters. Model specifications and choice of prior distributions match those described in Section \ref{sec:svbr_model}. As a comparison to previous studies, we additionally fit a fixed buffer model (i.e., SVBR$\left[5, \theta\right]$) where we choose the pre-specified radius of exposure as $\delta(\textbf{s}_j) = 5$ km for all $j$. We select a $5$ km radius based on the definition of accessible healthcare used in several previous studies \citep{Towongo2022}. For the other models, we define $\delta(\textbf{s}_j) \in (a = 0\text{ km}, b = 20\text{ km})$. This upper bound was chosen to be several times larger than the $5$ km threshold typically used to define accessible healthcare, allowing for significant flexibility in modeling the spatially-varying effects. 

We compare models using the previously described tools from Section \ref{sec:sim_study} (i.e., WAIC, PSIS-LOO, LPML). We also provide model validation checks via Bayesian p-values \citep{gelman1996posterior} and conditional predictive ordinates (CPOs) for each fitted model. Bayesian p-values are based on sampling replicate datasets from the posterior predictive distribution generated under the fitted model, constructing a test statistic as a function of the data and parameters, and determining how extreme the observed test statistic (i.e., using the real data) is with respect to those generated from the replicate data. Values near zero or one indicate that the fitted model is unable to capture important features of the observed dataset. We consider two test statistics that describe the general goodness-of-fit of the model to the data such that $$\text{TS}_1\left(\boldsymbol{Y}\right) = \frac{\sum_{j=1}^m \sum_{i=1}^{n_j} \left\{Y_i\left(\textbf{s}_j\right) - p_i\left(\textbf{s}_j\right)\right\}^2}{p_i\left(\textbf{s}_j\right)\left\{1 - p_i\left(\textbf{s}_j\right)\right\}}\ \text{and}$$ $$\text{TS}_2\left(\boldsymbol{Y}\right) = -2\sum_{j=1}^m \sum_{i=1}^{n_j} \left[Y_i\left(\textbf{s}_j\right) \ln\left\{p_i\left(\textbf{s}_j\right)\right\} + \left\{1 - Y_i\left(\textbf{s}_j\right)\right\}\ln\left\{1 - p_i\left(\textbf{s}_j\right)\right\}\right]$$ where $\boldsymbol{Y}$ is the complete vector of observed data and $\text{TS}_2\left(.\right)$ is based on the deviance of the Bernoulli likelihood. The Bayesian p-value for test statistic $j$ is estimated as $\frac{1}{S} \sum_{k=1}^S\mathbbold{1}\left\{\text{TS}_j\left(\boldsymbol{Y}_0^{(k)}\right) \geq \text{TS}_j\left(\boldsymbol{Y}\right)\right\}$ where $S$ is the number of collected posterior samples and $\boldsymbol{Y}_0^{\left(k\right)}$ is the replicate dataset simulated using the $k^{\text{th}}$ sample of model parameters from the joint posterior distribution. 

The CPO for $Y_i\left(\textbf{s}_j\right)$ is defined as $f\left(Y_i\left(\textbf{s}_j\right) | \boldsymbol{Y}_{-Y_i\left(\textbf{s}_j\right)}\right)$ where $\boldsymbol{Y}_{-Y_i\left(\textbf{s}_j\right)}$ is the complete vector of observed data with observation $Y_i\left(\textbf{s}_j\right)$ removed. It is estimated using results from \cite{Gelfand1993} for each observation in the dataset, with extremely small values indicating poor model fit for those particular observations.

\subsection{Results}
For each model and exposure definition, we collect a total of $10,000$ posterior samples from our MCMC algorithms, after discarding $50,000$ burn-in iterations and thinning by a factor of $20$ to reduce posterior autocorrelation. We assess convergence by visual inspection of trace plots and computing Geweke diagnostics \citep{geweke1991evaluating}  for key model parameters. Across all fitted models and monitored parameters, we observed no obvious signs of non-convergence.

We observe relatively consistent estimated associations between the maternal socio-demographic and cluster-level covariates included in the model and ANC usage across the different versions of SVBR (Table S11). Mothers with secondary or higher education and those who have stable employment are significantly more likely to complete $\ge 4$ ANC visits, while women of ``other" or ``no'' religion have significantly lower odds of receiving adequate care than Christians. Additionally, after adjustment for the individual-level factors, those mothers living in urban areas and, more specifically in the Menabe (vs.\ Androy) region, are less likely to receive adequate care.

Table \ref{table:application_results_waic_xadjusted_wadjusted} presents model fit and validation metrics for all competing methods and exposure definitions. The Bayesian p-values for both test statistics suggest that each of the proposed methods adequately describe the observed data in terms of general goodness-of-fit, as none of the values are near zero or one. In Figure S1, we present box plots of estimated CPO values are displayed for every fitted model with no obvious outliers observed.

Based on the findings of our simulation study (Section \ref{sec:sim_study}), WAIC, PSIS-LOO, and LPML are all useful metrics for understanding potential underlying data structures, including distinguishing between parameters that are constant versus spatially-varying. SVBR$\left[5, \theta\right]$ provides the poorest fit to the data, with improvements seen for models that allow for greater flexibility in the radii, even after accounting for increasing model complexity. Models that use counts of facilities to define ``exposure" generally outperform those that use the spherical exposure definition. Overall, the best model for explaining variability in ANC visits based on healthcare exposure (as indicated by every model comparison metric) is  SVBR$\left[\delta\left(\textbf{s}_j\right), \theta\right]$, with exposure defined as counts of facilities within the buffer radius. This suggests that, after accounting for variability in the radii, the exposure effect can be treated as constant across the domain.

\begin{figure}[h]
    \centering
    \includegraphics[width=\textwidth, trim = 0 1cm 0 0, clip]{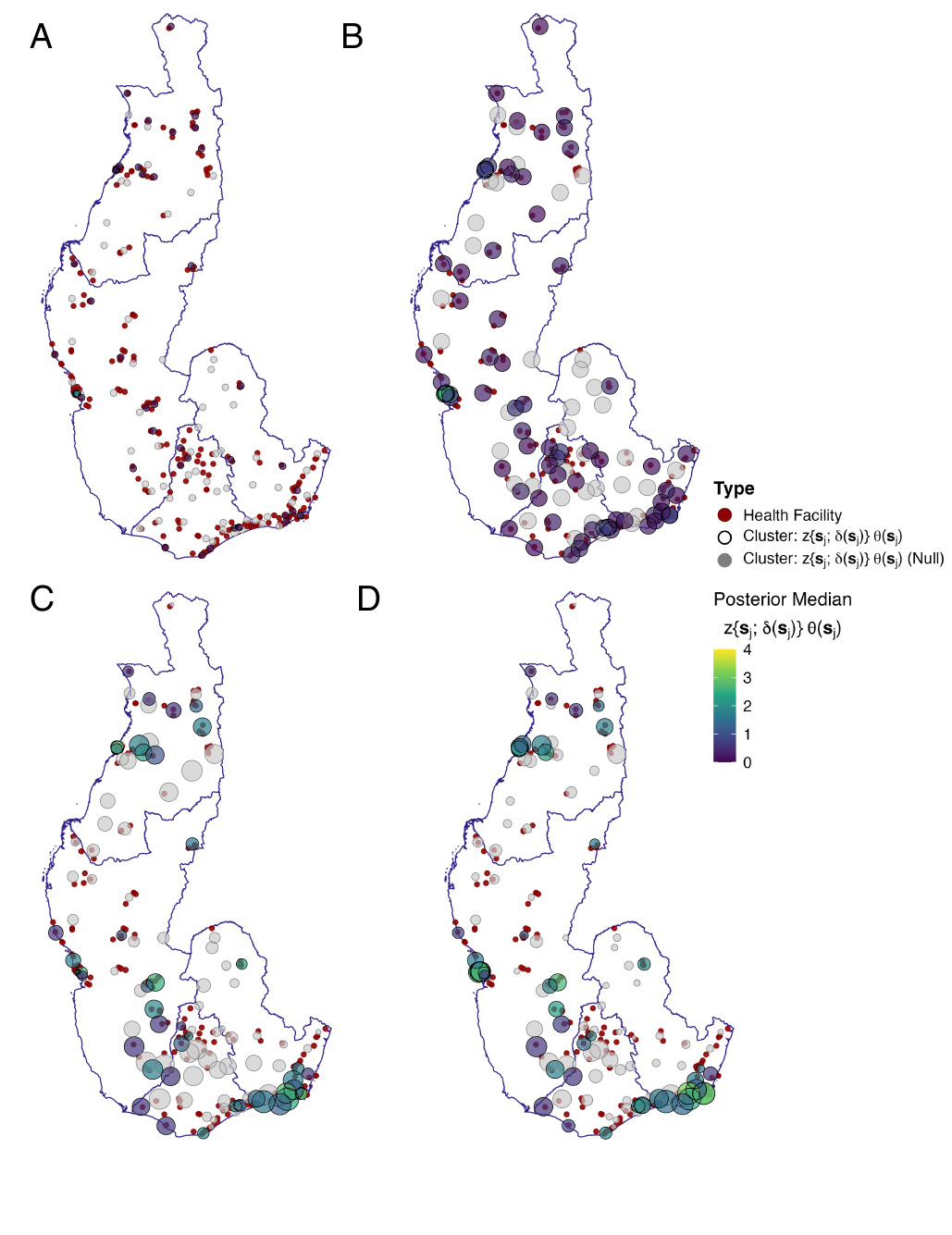}
    \caption{\small Results from the Madagascar (Toliara Province) antenatal care case study for the counts exposure definition. Posterior median radii estimates (transparent circles) are presented for each competing model ((A) SVBR$\left[5, \theta\right]$, (B) SVBR$\left[\delta, \theta\right]$, (C) SVBR$\left[\delta\left(\textbf{s}_j\right), \theta\right]$, (D) SVBR$\left[\delta\left(\textbf{s}_j\right), \theta\left(\textbf{s}_j\right)\right]$). Clusters where the $95\%$ highest posterior density interval for $\text{z}\{\textbf{s}_j; \delta(\textbf{s}_j)\} \theta(\textbf{s}_j)$ includes $0$ are indicated with grey shading while clusters whose interval does not include $0$ are shaded based on the corresponding posterior median. Health facility locations are identified with solid red points.}
    \label{fig:radii_maps_xadjusted_wadjusted_exposure0}
\end{figure}

Figure \ref{fig:radii_maps_xadjusted_wadjusted_exposure0} displays the spatial distribution of posterior median radii across the study area for the models using the counts exposure definition.  For SVBR$\left[5, \theta\right]$, an increase of one health facility within $5$ km is associated with a $16\%$ increase in the odds of completing at least $4$ ANC visits  ($\exp\left(\theta\right) = 1.16$; $95\%$ HDI: $(1.08, 1.25)$) (Figure \ref{fig:radii_maps_xadjusted_wadjusted_exposure0}A).  SVBR$\left[\delta, \theta\right]$ also includes a single radius for defining exposure, but allows the data to select the optimal distance. In this case, the radius was estimated as $12.33$ km ($95\%$ HDI: $(10.49, 15.01)$), with a similar exposure effect estimate as SVBR$\left[5, \theta\right]$ ($\exp\left(\theta\right) = 1.17$; $95\%$ HDI: $(1.11, 1.24)$) (Figure \ref{fig:radii_maps_xadjusted_wadjusted_exposure0}B). Both models allow for a straightforward interpretation of a single radius and exposure effect parameter, but the poor relative performance of SVBR$\left[5, \theta\right]$ and SVBR$\left[\delta, \theta\right]$ suggest that these models fail to adequately describe the variability in the data. 

Results from SVBR$\left[\delta\left(\textbf{s}_j\right), \theta\right]$ suggest substantial spatial variability in the radii, as indicated by the improved model comparison scores. Each additional healthcare facility within an individual's exposure buffer increases the odds of completing $\ge4$ ANC visits by $86\%$ ($\exp\left(\theta\right) = 1.86$; $95\%$ HDI: $(1.57, 2.18)$)  (Figure \ref{fig:radii_maps_xadjusted_wadjusted_exposure0}C). In Table S11, results from the buffer radii regression model suggest that as GHFI increases (i.e., an area is more urban), the buffer radii shrink, with no other covariates having a significant impact.

SVBR$\left[\delta\left(\textbf{s}_j\right), \theta(\textbf{s}_j)\right]$ allows for both spatial variability in the radii and exposure effects. However, as indicated by the model comparison results, there are not substantial differences between results from this model and those from SVBR$\left[\delta\left(\textbf{s}_j\right), \theta\right]$ (Figure \ref{fig:radii_maps_xadjusted_wadjusted_exposure0}D). The covariate results from the exposure effect regression model (Table S11) do indicate effect modification is present. The 95\% HDI for the interaction between exposure and GHFI is entirely below zero, indicating that clusters in areas with higher GHFI may have a smaller effect size on average. However, given the high degree of similarity between the results in Figure \ref{fig:radii_maps_xadjusted_wadjusted_exposure0}, and the model comparison findings, we select SVBR$\left[\delta\left(\textbf{s}_j\right), \theta\right]$ as the primary model for statistical inference.

\begin{figure}[h]
    \centering
        \centering
        \includegraphics[width=0.8\textwidth]{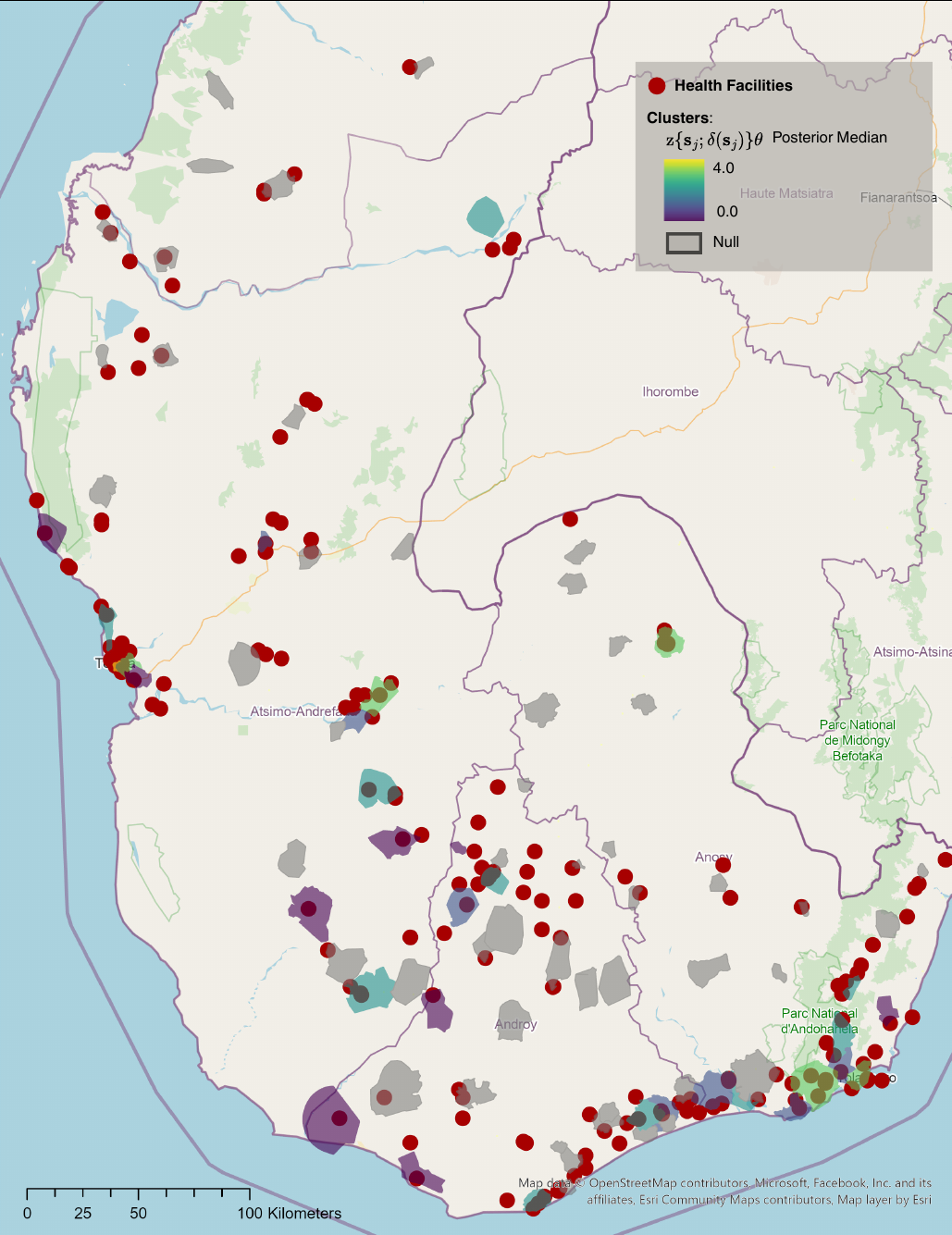}
    \caption{Distance buffer polygons mapping the cluster-level posterior median radius on walking road/path networks, where health facility exposure is defined as counts of facilities within distance $\delta(\textbf{s}_j)$. Buffers are colored based on the posterior median estimate of $\text{z}\{\textbf{s}_j; \delta(\textbf{s}_j)\}\theta$ obtained from applying SVBR$\left[\delta\left(\textbf{s}_j\right), \theta\right]$ to the Madagascar (Toliara Province) case study data. The map shows a subset of the larger study region for improved visibility.}
    \label{fig:gis_results_map}
\end{figure}

While in Figure \ref{fig:radii_maps_xadjusted_wadjusted_exposure0} we map the buffer radii and simple straight line buffers for visualization purposes, in reality our underlying exposure is based on walking distances on path/road networks. Thus, Figure \ref{fig:gis_results_map} displays the actual buffer shapes of the posterior median radii from the primary model on paths and road networks for select clusters.  

Across all methods there is a clear association between access to health facilities and completing at least four ANC visits. There are also clear signs of spatial variability in these associations across the Toliara region of Madagascar, leading to differences in model fit and interpretation between the competing approaches, with SVBR$\left[\delta\left(\textbf{s}_j\right), \theta\right]$ providing the best combination of fit and complexity overall.   

\subsection{Sensitivity analyses}\label{sec:sens}
We carry out two sensitivity analyses to test whether jointly modeling the buffer radii and exposure effect parameter(s) affects parameter identifiability in SVBR$\left[\delta\left(\textbf{s}_j\right), \theta\right]$ and SVBR$\left[\delta\left(\textbf{s}_j\right), \theta\left(\textbf{s}_j\right)\right]$. We begin by calculating the posterior median buffer radii from both fitted models. Using these values, we calculate the exposure for every cluster location and therefore, individual. We then fit standard non-Bayesian logistic regression models using these exposures as a known covariate to estimate $\boldsymbol{\eta}$ (i.e., parameters that define $\theta$, $\theta(\textbf{s}_j)$). For SVBR$\left[\delta\left(\textbf{s}_j\right), \theta\right]$, this model includes the exposure and $\textbf{x}_i\left(\textbf{s}_j\right)$ covariates. For SVBR$\left[\delta\left(\textbf{s}_j\right), \theta(\textbf{s}_j)\right]$, the model includes the exposure, $\textbf{x}_i\left(\textbf{s}_j\right)$, and interactions between exposure and each entry of $\textbf{q}\left(\textbf{s}_j\right)$. If the SVBR models are performing as expected and are identifiable, we anticipate similar $\boldsymbol{\eta}$ and $\boldsymbol{\beta}$ estimates will be produced for these simplified models. 

For SVBR$\left[\delta\left(\textbf{s}_j\right), \theta\right]$, the simplified model's estimate of $\exp\left\{\theta\right\}$ is $1.90$ (95\% confidence interval (CI): $1.73$, $2.09$) compared to $1.86$ (95\% HDI: $1.57$, $2.18$). The point estimate matches closely while the uncertainty is reduced in the simplified model due to treating the buffer radii as known. The findings are similar for SVBR$\left[\delta\left(\textbf{s}_j\right), \theta(\textbf{s}_j)\right]$. In Figure S2, we present scatter plots of the $\boldsymbol{\eta}$ and $\boldsymbol{\beta}$ point estimates for both sensitivity analyses, further showing strong agreement between the two models. Overall, the results demonstrate that the model parameters are identifiable even when buffer radii and exposure effects are estimated simultaneously in SVBR.

In the final sensitivity analysis, we assess the robustness of our covariate-adjustment approach for the DHS survey design by comparing it to a traditional survey-weighted analysis. We fit a survey-weighted logistic regression model using the \textbf{svyglm} function in R \citep{survey} with exposures defined by a fixed buffer radius of $12.33$ km (the posterior median from SVBR$\left[\delta, \theta\right]$), the same covariates, and DHS-provided sampling weights. We compare these results to those from SVBR$[12.33, \theta]$ (Table S12). The exposure effect estimates are nearly identical: $1.16$ ($95\%$ CI: $1.09, 1.23$) for the survey-weighted model versus $1.18$ ($95\%$ HDI: $1.12, 1.24$) for our Bayesian covariate-adjustment approach. Individual-level covariate estimates also showed strong agreement. These findings suggest our results are robust to the survey design modeling approach.

\section{Discussion}\label{sec:discussion}
In this work, we developed SVBR, a hierarchical Bayesian spatial change points framework for place-based epidemiology studies that jointly estimates the spatial range of exposure effects (buffer radii) and the magnitude of the exposure/outcome relationship.  SVBR$\left[\delta\left(\textbf{s}_j\right), \theta(\textbf{s}_j)\right]$ accommodates spatial heterogeneity in radii and exposure effects, accounts for spatial correlation, and offers a flexible framework suitable for diverse data types and epidemiology modeling settings. Through realistic simulations grounded in real-world data, we evaluated different versions of SVBR across four distinct simulation settings. Our results show that the SVBR$\left[\delta\left(\textbf{s}_j\right), \theta(\textbf{s}_j)\right]$ maintains strong performance across both simple and complex spatial structures. It effectively captures true variation in radii and exposure effects, but also adapts to simpler settings without spatial variability in parameters. In contrast, models that assume a single buffer radius and exposure effect (i.e., SVBR$\left[\delta, \theta\right]$) are not adequate in settings where there is true spatial variability. 

We also apply SVBR to investigate the relationship between walking distance to healthcare facilities and odds of completing at least four ANC visits during pregnancy in the Toliara Province of Madagascar. Understanding how distance affects healthcare access, utilization, and outcomes is of growing interest in public health literature, but many studies assume fixed buffer radii and homogeneous effects \citep{Hierink2021}. Our results reveal substantial spatial variability in both the effective range of healthcare access and its impact on ANC utilization. For SVBR$\left[\delta\left(\textbf{s}_j\right), \theta\right]$, we find that the association between facility proximity and receiving care varies considerably across the region, with posterior median buffer radii ranging from $2.47$ ($95\% \text{ HDI: } 1.86, 3.30$) km to $17.08$ ($95\% \text{ HDI: } 12.64, 19.01$) km across clusters, and urbanized areas (higher GHFI) having smaller buffer radii on average.

These findings have important methodological and policy implications. The substantial improvement in model fit when allowing spatially-varying radii demonstrates that uniform distance-based policies may inadequately capture geographic heterogeneity in healthcare access. Our results suggest that a single definition of accessible care, such as the widely used $5$ km cutoff, may fail to capture accessibility in heterogeneous spatial contexts. A single distance threshold will misclassify accessibility for many communities in settings like Toliara Province, where effective access distances vary substantially. Our results suggest that facility placement strategies, accessibility standards, and resource allocation decisions should be context-specific rather than applying uniform distance criteria. For example, mobile clinic routing or new facility siting in rural areas might need to account for access distances exceeding $15$ km, whereas in more urban areas, distances under $10$ km may be sufficient. More broadly, while proximity to facilities is a modifiable determinant of maternal health outcomes, the spatial variation in this relationship indicates that simply increasing facility density may not have uniform effects, and targeted interventions should consider local patterns of access and utilization.

The SVBR framework allows for varying degrees of complexity, from SVBR$\left[\delta, \theta\right]$ to models that allow for variability in both radii and exposure effects. While traditional single-buffer models offer simplicity in terms of implementation and interpretation, they may miss important spatial patterns. In contrast, flexible models like SVBR balance complexity and interpretability, and offer a more nuanced perspective on how the exposure/effect relationship varies across space. Notably, all models within the SVBR framework, regardless of specification, allow for direct inference on the buffer radii.

While SVBR flexibly captures spatial variation, further extensions are needed to handle more complex spatial interactions and dependencies, including overlapping exposure zones and context-specific interference or synergy in effects. For example, it might be sensible to model competing effects in healthcare access where overlapping service areas might increase care availability, but not in pollutant exposures where overlaps may not alter individual risk. Tailoring models to specific exposure contexts will further improve inference. While SVBR is agnostic to the exposure being studied, the current implementation provides a flexible framework for future researchers to adapt models to their specific area of application.

Many epidemiologic studies examine geo-referenced exposures, but few methods allow for flexible estimation of their spatial extent and effect magnitude. SVBR addresses this gap, providing a framework for estimating buffer radii and associated exposure effects. Overall, SVBR offers a promising, flexible, and data-driven approach for investigating spatially varying exposure-outcome relationships in place-based epidemiology studies.  The \textbf{EpiBuffer} R package (\url{https://github.com/warrenjl/EpiBuffer}) can facilitate future applied use and methodological extensions in this area. 

\begin{table}
\centering
\caption{Description of the study sample ($n=1,818$) by antenatal care (ANC) status for the Madagascar (Toliara Province) case study. For categorical variables, counts and proportions are reported. For continuous variables, means and standard deviations are reported.}\label{table:sample_covariates}
\centering
\footnotesize
\begin{tabular}[t]{lrr}
\toprule
\multicolumn{1}{c}{ } & \multicolumn{1}{c}{ANC $<4$ ($n = 883$)} & \multicolumn{1}{c}{ANC $\ge 4$ ($n = 935$)} \\
\midrule
\textbf{Individual-level Characteristics}  &  &   \\
Age (Years) &  &   \\

\hspace{1em}[35-50) & 166 (18.80) & 143 (15.29)\\

\hspace{1em}[20-35) & 520 (58.89) & 590 (63.10)\\

\hspace{1em}[12-20) & 197 (22.31) & 202 (21.60)\\

Education &  &   \\

\hspace{1em}None & 481 (54.47) & 344 (36.79)\\

\hspace{1em}Primary & 315 (35.67) & 335 (35.83)\\

\hspace{1em}Secondary or Higher & 87 (9.85) & 256 (27.38)\\

Employment Status &  &   \\

\hspace{1em}No & 153 (17.33) & 162 (17.33)\\

\hspace{1em}Yes & 730 (82.67) & 773 (82.67)\\

Marital Status &  &   \\

\hspace{1em}Married/Cohabitating & 621 (70.33) & 668 (71.44)\\

\hspace{1em}Never/Widowed/Divorced/Separated & 262 (29.67) & 267 (28.56)\\

Parity &  &   \\

\hspace{1em}First & 174 (19.71) & 227 (24.28)\\

\hspace{1em}Not First & 709 (80.29) & 708 (75.72)\\

Religion &  &   \\

\hspace{1em}Christian & 308 (34.88) & 527 (56.36)\\

\hspace{1em}None & 522 (59.12) & 382 (40.86)\\

\hspace{1em}Other Religion & 53 (6.00) & 26 (2.78)\\

\textbf{Cluster-level Characteristics}  &  &   \\
Residence &  &   \\

\hspace{1em}Rural & 763 (86.41) & 780 (83.42)\\

\hspace{1em}Urban & 120 (13.59) & 155 (16.58)\\

Region & & \\
\hspace{1em} Androy & 226 (25.59) & 274 (29.30) \\
\hspace{1em} Anosy & 225 (25.48) & 220 (23.53) \\
\hspace{1em} Atsimo-Andrefana & 195 (22.08) & 257 (27.49) \\
\hspace{1em} Menabe & 237 (26.84) & 184 (19.68) \\

Elevation (Meters) & 279.16 (253.74) & 209.82 (207.36)\\

Wealth (Proportion in Upper Quintiles) &  0.10 (0.22) &  0.24 (0.32) \\

Global Human Footprint Index ($0-100$) & 25.14 (8.74) & 29.27 (10.03)\\

\bottomrule
\end{tabular}
\end{table}

\clearpage 

\begin{table}
\centering
\caption{Spatially-Varying Buffer Radii (SVBR) model descriptions and notation.}\label{table:model_desc}
\begin{tabular}[t]{lll}
\toprule
Model & Radius ($\delta$) & Exposure Effect ($\theta$) \\
\midrule
SVBR$\left[\text{x}, \theta\right]$ & One value fixed at $\text{x}$ & One value, estimated \\
SVBR$\left[\delta, \theta\right]$ & One value, estimated & One value, estimated \\
SVBR$\left[\delta\left(\textbf{s}_j\right), \theta\right]$ & Spatially-varying, estimated & One value, estimated \\
SVBR$\left[\delta\left(\textbf{s}_j\right), \theta\left(\textbf{s}_j\right)\right]$ & Spatially-varying, estimated & Spatially-varying, estimated \\
\bottomrule
\end{tabular}
\end{table}
\clearpage

\begin{table}
\centering
\caption{Mean absolute error (MAE) of posterior median estimates in the simulation study with Bernoulli outcome. For each dataset, MAE is computed for each parameter and averaged across spatial locations when applicable; these values are then averaged across all $500$ simulated datasets. Standard errors are given in parentheses. Bolded values indicate the smallest MAE in each row.}\label{table:sim_mse}
\centering
\begin{tabular}[t]{lrrr}
\toprule
\multicolumn{1}{c}{ } & \multicolumn{3}{c}{Model Type} \\
\cmidrule(l{3pt}r{3pt}){2-4}
Parameter & SVBR$\left[\delta, \theta\right]$ & SVBR$\left[\delta\left(\textbf{s}_j\right), \theta\right]$ & SVBR$\left[\delta\left(\textbf{s}_j\right), \theta\left(\textbf{s}_j\right)\right]$\\
\midrule
\addlinespace[0.3em]
\multicolumn{4}{l}{\textbf{No Effect}}\\
\hspace{1em}$\text{z}\left(\textbf{s}_j; \delta\right)\theta$ & \textbf{0.02} (0.00) & 0.04 (0.00) & 0.04 (0.00)\\
\hspace{1em}$\beta_1$ & \textbf{0.21} (0.01) & 0.21 (0.01) & 0.21 (0.01)\\
\addlinespace[0.3em]
\multicolumn{4}{l}{\textbf{Single Radius \& Effect}}\\
\hspace{1em}$\delta$ & \textbf{0.51} (0.03) & 1.01 (0.03) & 0.99 (0.03)\\
\hspace{1em}$\theta$ & \textbf{0.08} (0.00) & 0.09 (0.00) & 0.11 (0.00)\\
\hspace{1em}$\text{z}\left(\textbf{s}_j; \delta\right) \theta$ & \textbf{0.18} (0.01) & 0.22 (0.01) & 0.29 (0.01)\\
\hspace{1em}$\beta_1$ & \textbf{0.23} (0.01) & 0.24 (0.01) & 0.24 (0.01)\\
\addlinespace[0.3em]
\multicolumn{4}{l}{\textbf{Varying Radii, Single Effect}}\\
\hspace{1em}$\delta(\textbf{s}_j)$ & 5.81 (0.09) & \textbf{3.81} (0.04) & 3.83 (0.04)\\
\hspace{1em}$\theta$ & 0.53 (0.10) & \textbf{0.14} (0.01) & 0.17 (0.01)\\
\hspace{1em}$\text{z}\left\{\textbf{s}_j; \delta(\textbf{s}_j)\right\}\theta$ & 0.64 (0.01) & \textbf{0.41} (0.01) & 0.45 (0.01)\\
\hspace{1em}$\beta_1$ & \textbf{0.23} (0.01) & 0.23 (0.01) & 0.23 (0.01)\\
\hspace{1em}$\gamma_1$ & -- & \textbf{0.18} (0.01) & 0.18 (0.01)\\
\addlinespace[0.3em]
\multicolumn{4}{l}{\textbf{Varying Radii \& Effect}}\\
\hspace{1em}$\delta(\textbf{s}_j)$ & 6.43 (0.10) & 3.98 (0.04) & \textbf{3.81} (0.04)\\
\hspace{1em}$\theta(\textbf{s}_j)$ & 1.00 (0.16) & 0.21 (0.01) & \textbf{0.18} (0.01)\\
\hspace{1em}$\text{z}\left\{\textbf{s}_j; \delta(\textbf{s}_j)\right\} \theta(\textbf{s}_j)$ & 0.73 (0.01) & 0.44 (0.01) & \textbf{0.41} (0.01)\\
\hspace{1em}$\beta_1$ & \textbf{0.23} (0.01) & 0.24 (0.01) & 0.24 (0.01)\\
\hspace{1em}$\gamma_1$ & -- & 0.26 (0.01) & \textbf{0.21} (0.01)\\
\hspace{1em}$\eta_1$ & -- & -- & \textbf{0.12} (0.01)\\
\addlinespace[0.3em]
\multicolumn{4}{l}{\textbf{Misspecified}}\\
\hspace{1em}$\delta(\textbf{s}_j)$ & 3.12 (0.05) & 2.43 (0.03) & \textbf{2.39} (0.03)\\
\hspace{1em}$\theta(\textbf{s}_j)$ & 0.17 (0.01) & \textbf{0.14} (0.00) & 0.16 (0.00)\\
\hspace{1em}$\text{z}\left\{\textbf{s}_j; \delta(\textbf{s}_j)\right\} \theta(\textbf{s}_j)$ & 0.39 (0.01) & \textbf{0.31} (0.00) & 0.36 (0.01)\\
\hspace{1em}$\beta_1$ & \textbf{0.21} (0.01) & 0.21 (0.01) & 0.22 (0.01)\\
\hspace{1em}$\gamma_1$ & -- & 0.25 (0.01) & \textbf{0.19} (0.01)\\
\bottomrule
\end{tabular}
\end{table}
\clearpage 

\begin{landscape}
\begin{table}
\centering
\caption{Model comparison and validation results for the Madagascar (Toliara Province) antenatal care case study. For model comparisons, Watanabe-Akaike information criteria (WAIC), Pareto-smoothed importance sampling leave-one-out (PSIS-LOO) cross validation, and logarithm of the pseudo marginal likelihood (LPML) are displayed. Each metric is transformed so that smaller values indicate improved model fit.  The effective number of parameters, an estimate of model complexity, is given in parentheses for WAIC (p$_{\text{WAIC}}$) and PSIS-LOO (p$_{\text{LOO}}$).  The bolded value indicates the smallest value across all competing approaches. For model validation, the estimated Bayesian p-values for the two test statistics detailed in Section \ref{sec:application_study} are displayed.}\label{table:application_results_waic_xadjusted_wadjusted}
\centering
\begin{tabular}[t]{llrrrrr}
\toprule
Model                                                                               & Exposure       & WAIC (p$_{\text{WAIC}}$) & PSIS-LOO (p$_{\text{LOO}}$)   & LPML & TS$_1\left(\boldsymbol{Y}\right)$ & TS$_2\left(\boldsymbol{Y}\right)$ \\
\hline
SVBR$\left[5, \theta\right]$                                                        & Count     & 2353.00 (15.56)          & 2353.02 (15.57)          & 2353.00  & 0.17 & 0.22 \\
                                                                                    & Spherical & 2341.87 (15.33)          & 2341.89 (15.34)          & 2341.87 & 0.21 & 0.23 \\
SVBR$\left[\delta, \theta\right]$                                                   & Count     & 2324.65 (16.66)          & 2324.71 (16.70)          & 2324.69 & 0.17 & 0.23 \\
                                                                                    & Spherical & 2329.91 (16.13)          & 2329.94 (16.14)          & 2329.92 & 0.11 & 0.23 \\
SVBR$\left[\delta\left(\textbf{s}_j\right), \theta\right]$                          & Count     & \textbf{2214.72} (52.65) & \textbf{2215.18} (52.88) & \textbf{2215.12} & 0.25 & 0.29 \\
                                                                                    & Spherical & 2306.35 (19.31)          & 2306.38 (19.33)          & 2306.36 & 0.28 & 0.23 \\
SVBR$\left[\delta\left(\textbf{s}_j\right), \theta(\textbf{s}_j)\right]$ & Count     & 2218.14 (53.78)          & 2218.89 (54.16)          & 2218.78 & 0.21 & 0.23 \\
                                                                                    & Spherical & 2255.16 (38.54)          & 2255.32 (38.63)          & 2255.28 & 0.22 & 0.22 \\
\bottomrule
\end{tabular}
\end{table}
\end{landscape}
\clearpage 

\clearpage

\section*{Funding}
This work was partially funded by the National Institute on Drug Abuse (R01 DA060716) and the National Institute of Environmental Health Sciences (R01 ES028346; R01 ES034756).

\section*{Data availability}
The health care facility data that support the findings of this study are openly available in the Springer Nature figshare repository at \url{https://doi.org/10.6084/m9.figshare.7725374.v1}. The DHS cluster data are available from The Demographic and Health Surveys Program. Restrictions apply to the availability of these data, which were used under permission from the DHS program for the purposes of this study. Data are available at \url{https://dhsprogram.com/data/dataset/Madagascar_Standard-DHS_2021.cfm?flag=0} with the permission of the DHS program.

\clearpage
\appendix

\begin{titlepage}
  \centering
  {\Large\bfseries Supplementary Materials \par}
  \vspace{1.5em}
  {\large Modeling Spatial Heterogeneity in Exposure Buffers and Risk: \\
  A Hierarchical Bayesian Approach \par}
  \vspace{2em}
  {\large Saskia Comess\textsuperscript{1}, Daniel E. Ho\textsuperscript{2}, Joshua L. Warren\textsuperscript{3*} \par}
  \vspace{1em}
  {\small
  \textsuperscript{1}{Emmett Interdisciplinary Program in Environment and Resources, Stanford University, CA, USA \\
  \textsuperscript{2}Regulation, Evaluation and Governance Lab, Stanford Law School, CA, USA \\
  \textsuperscript{3}Department of Biostatistics, Yale University, CT, USA \\
  \textsuperscript{*}Corresponding author: joshua.warren@yale.edu
  }}
  \vfill
\end{titlepage}

\section{SVBR model fitting details and derivations}

\subsection{$K << m$ approximation for computational efficiency}\label{sec:computational_approximation}

With the standard implementation of SVBR (Spatially-Varying Buffer Radii), we directly model the spatial random effects $\phi(\textbf{s}_j)$ for each location $j = 1,\hdots,m$. With a large number of unique geo-referenced locations, fitting this model will be computationally burdensome or even infeasible \citep{Heaton2019}. To improve the computational efficiency of the model in large $m$ scenarios, we implement the dimensionality reducing predictive process model approach described by \citeauthor{Banerjee2008} (\citeyear{Banerjee2008}, \citeyear{Banerjee2012}). In brief, this model assumes that the spatial information available from the entire set of observed $m$ locations can be summarized in terms of a smaller, but representative, fixed set of $K$ locations (``knots'') ($K << m$) using a predictive process, $\{\tilde{\phi}(\textbf{s}_j), j = 1,\hdots,m\}$. The knot locations $\{\textbf{s}_1^*,\hdots,\textbf{s}_K^*\}$ may or may not form a subset of the entire collection of observed locations $\{\textbf{s}_1,\hdots,\textbf{s}_m\}$. 

The parent process for the spatial random effects follows a zero-centered Gaussian process with exponential covariance function $\tau_{\phi}^2 \Sigma(\rho_{\phi})_{j,j'} = \tau_{\phi}^2 \exp\{-\rho_{\phi} || \textbf{s}_j-\textbf{s}_{j'}||\}$, as described in Section \ref{sec:svbr_radii_theta_specifications} of the main text. The corresponding predictive process replaces $\phi\left(\textbf{s}_j\right)$ with
    \begin{equation*}\label{eq:phi_tilde}
         \tilde{\phi}(\textbf{s}_j) = \textbf{c}(\textbf{s}_j; \rho_{\phi})^{\text{T}} \Sigma^*(\rho_{\phi})^{-1} \boldsymbol{\phi^*}\\
    \end{equation*} 
where for each spatial location, $\textbf{c}(\textbf{s}_j; \rho_{\phi})^{\text{T}}$ is a $1 \times K$ vector describing the correlation between parameters at the observed location $\textbf{s}_j$ and each knot location $\textbf{s}_k^*$ such that 
$$\textbf{c}(\textbf{s}_j; \rho_{\phi})^{\text{T}} = \begin{bmatrix}\exp\left\{-\rho_{\phi} || \textbf{s}_j - \textbf{s}_1^*||\right\} & \hdots & \exp\left\{-\rho_{\phi} || \textbf{s}_j - \textbf{s}_K^*||\right\}\end{bmatrix}.$$ 
The $K \times K$ matrix $\Sigma^*(\rho_{\phi})$ specifies the correlation between the knot locations, with elements $\Sigma_{k, k'}^* = c(\textbf{s}_{k}^*, \textbf{s}_{k'}^*; \rho_{\phi}) = \exp\left\{-\rho_{\phi} || \textbf{s}_{k}^* - \textbf{s}_{k'}^*||\right\}$. The complete vector of spatial random effects across all knot locations, $\boldsymbol{\phi}^{*\text{T}} =  \left\{\phi^*\left(\textbf{s}^*_1\right),\hdots,\phi^*\left(\textbf{s}^*_K\right)\right\}$, also follows a zero-centered Gaussian process with correlation matrix $\Sigma^*(\rho_{\phi})$ such that $\boldsymbol{\phi}^* | \tau_{\phi}^2, \rho_{\phi} \sim \text{MVN}\{\boldsymbol{0}_K, \tau_{\phi}^2 \Sigma^*(\rho_{\phi})\}$.
The predictive process model only impacts the specification of the spatial random effect $\phi(\textbf{s}_j)$, replacing it with $\tilde{\phi}(\textbf{s}_j)$ in (\ref{eq:spatial_radii}) from the main text. The rest of the model specification remains unchanged. 

This approximation is available for use in the \textbf{EpiBuffer} R package to facilitate future applications that include a large number of unique spatial locations, but was not used in any of the analyses presented in the main text. 
 
\subsection{SVBR in matrix form and other notation}\label{svbr_supplement_derivations}
SVBR follows the specification provided in (\ref{eq:svbr}), (\ref{eq:spatial_radii}), and (\ref{eq:theta_polynomial}) from the main text. For the purposes of deriving the full conditional distributions needed for posterior sampling, we rewrite the main model and sub-models in matrix form.

\begin{itemize}
    \item The regression component of SVBR can be expressed in matrix form such as: 

    \begin{equation*}
    \begin{aligned}
    g\left(\boldsymbol{\mu}\right) &= \textbf{O} + \textbf{X} \boldsymbol{\beta} + \widetilde{\textbf{Z}}\left(\boldsymbol{\delta}\right)\boldsymbol{\eta}, \\[1.5ex]
    \end{aligned}
    \end{equation*}

    \begin{itemize}
    
    \item $g\left(.\right)$:  selected link function based on likelihood choice
        
        \item $\boldsymbol{\mu}^{\text{T}} = \left\{\mu_1\left(\textbf{s}_1\right), \hdots, \mu_{n_1}\left(\textbf{s}_1\right), \mu_1\left(\textbf{s}_2\right), \hdots, \mu_{n_m}\left(\textbf{s}_m\right)\right\}$: $\sum_{j=1}^m n_j$ length vector 
        
        \item $\textbf{O}^{\text{T}} = \left\{\text{O}_1\left(\textbf{s}_1\right), \hdots, \text{O}_{n_1}\left(\textbf{s}_1\right), \text{O}_1\left(\textbf{s}_2\right), \hdots,  \text{O}_{n_m}\left(\textbf{s}_m\right)\right\}$: $\sum_{j=1}^m n_j$ length vector of offset terms (optional) for each individual $i$ at location $\textbf{s}_j$
        
        \item $\textbf{X} = 
                \begin{bmatrix}
                \textbf{x}_1(\textbf{s}_1)^\text{T} \\
                \vdots\\
                \textbf{x}_{n_1}(\textbf{s}_1)^\text{T} \\
                \textbf{x}_1(\textbf{s}_2)^\text{T} \\
                \vdots \\
                \textbf{x}_{n_m}(\textbf{s}_{m})^\text{T}
                \end{bmatrix}$: $\sum_{j=1}^m n_j \times p_x$ matrix of individual-specific covariates (includes an intercept)
        
        \item $\boldsymbol{\beta}^{\text{T}} = \left(\beta_0, \hdots, \beta_{p_x}\right)$: $p_x$ length vector of regression coefficients 
       
        \item $\boldsymbol{\delta}^{\text{T}} = \{\delta\left(\textbf{s}_1\right),\hdots,\delta\left(\textbf{s}_m\right)\}$: Location-specific radii
        
        \item $\widetilde{\textbf{Z}}\left(\boldsymbol{\delta}\right) = \begin{bmatrix} 
       \text{z}\{\textbf{s}_1; \delta(\textbf{s}_1)\}\text{q}_1\left(\textbf{s}_1\right)\boldsymbol{1}_{n_1} & \hdots & \text{z}\{\textbf{s}_1; \delta(\textbf{s}_1)\}\text{q}_{p_q}\left(\textbf{s}_1\right)\boldsymbol{1}_{n_1} \\
        \vdots\\
        \text{z}\{\textbf{s}_m; \delta(\textbf{s}_m)\}\text{q}_1\left(\textbf{s}_m\right)\boldsymbol{1}_{n_m} & \hdots & \text{z}\{\textbf{s}_m; \delta(\textbf{s}_m)\}\text{q}_{p_q}\left(\textbf{s}_m\right)\boldsymbol{1}_{n_m}\end{bmatrix}$: $\sum_{j=1}^m n_j \times p_q$ exposure matrix, scaled by the location-specific covariates contained in the $p_q$ length $\textbf{q}(\textbf{s}_j)$ vector (includes an intercept), where $\text{z}\left\{\textbf{s}_j; \delta(\textbf{s}_j)\right\}$ is the exposure for individuals at location $\textbf{s}_j$ based on buffer radius $\delta(\textbf{s}_j)$ and $\boldsymbol{1}_n$ is a length $n$ column vector of ones. All outcome units $i$ at the same location $\textbf{s}_j$ have the same exposure value. 
        
        \item $\boldsymbol{\eta}^{\text{T}} = \left(\eta_0, \hdots, \eta_{p_q - 1}\right)^{\text{T}}$: $p_q$ length vector of regression parameters for the location-specific predictors
    \end{itemize}
\end{itemize}

\begin{itemize}
    \item The spatially-varying radii model (\ref{eq:spatial_radii}) is specified in matrix form as: 
    \begin{equation*}
        \Phi^{-1} \left\{ \frac{\boldsymbol{\delta} - a}{b - a}\right\} = \textbf{W} \boldsymbol{\gamma} + \boldsymbol{\phi}
    \end{equation*}

    \begin{itemize}
        \item $\textbf{W} = \begin{bmatrix}
        \textbf{w}\left(\textbf{s}_1\right)^\text{T}\\
        \vdots \\
        \textbf{w}\left(\textbf{s}_m\right)^\text{T}
        \end{bmatrix}$: $m \times p_w$ matrix of location-specific predictors (includes an intercept)
        
        \item $\boldsymbol{\gamma}^{\text{T}} = \left(\gamma_0, \hdots, \gamma_{p_w - 1}\right)$: $p_w$ length vector of regression coefficients for the location-specific predictors
        
        \item $\boldsymbol{\phi}^{\text{T}} = \left\{\phi\left(\textbf{s}_1\right), \hdots, \phi\left(\textbf{s}_m\right)\right\}$: $m$ length vector of spatial random effects
    \end{itemize}
\end{itemize}

\subsection{SVBR posterior sampling}
\subsubsection{Choice of likelihood}
Updating of most parameters within the Markov chain Monte Carlo algorithm is straightforward (i.e., they have a standard, closed-form full conditional distribution) for multiple likelihood choices that cover a number of relevant outcome data types, including Gaussian with identity link function (continuous outcome), Bernoulli or binomial with logit link function (binary outcome), and negative binomial with logit link function (count data). The latter two likelihood/link function results are derived using the results from \citeauthor{polson2013bayesian} (2013), described in further detail in the following section. 

\begin{itemize}
    \item For Gaussian data with identity link function we have 
    \begin{equation*}
        Y_i(\textbf{s}_j) = \textbf{x}_i(\textbf{s}_j)^{\text{T}}\boldsymbol{\beta} + \text{z}\{\textbf{s}_j; \delta(\textbf{s}_j)\} \theta\left(\textbf{s}_j\right) + \epsilon_i,\ \epsilon_i|\sigma^2_{\epsilon} \stackrel{\text{iid}}{\sim}\text{N}\left(0, \sigma^2_{\epsilon}\right) 
    \end{equation*} 

    \item For binomial data with logit link function we have 
    \begin{equation*}
    \begin{aligned}
        Y_i(\textbf{s}_j) | p_i(\textbf{s}_j) &\stackrel{\text{ind}}{\sim} \text{Binomial}\left\{\tilde{n}_i(\textbf{s}_j), p_i(\textbf{s}_j)\right\},\\
        \ln\left\{\frac{p_i(\textbf{s}_j)}{1 - p_i(\textbf{s}_j)}\right\} &= \textbf{x}_i(\textbf{s}_j)^{\text{T}}\boldsymbol{\beta} + \text{z}\{\textbf{s}_j; \delta(\textbf{s}_j)\} \theta\left(\textbf{s}_j\right)
    \end{aligned}
    \end{equation*}  
    where $\tilde{n}_i(\textbf{s}_j)$ is the number of trials for outcome unit $i$ at location $\textbf{s}_j$.

    \item For Negative Binomial data with logit link function we have 
    \begin{equation*}
    \begin{aligned}
        Y_i(\textbf{s}_j) | r, p_i(\textbf{s}_j) &\stackrel{\text{ind}}{\sim} \text{Negative Binomial}\left\{r, p_i(\textbf{s}_j)\right\},\\
        \ln\left\{\frac{p_i(\textbf{s}_j)}{1 - p_i(\textbf{s}_j)}\right\} &= \text{O}_i(\textbf{s}_j) + \textbf{x}_i(\textbf{s}_j)^{\text{T}}\boldsymbol{\beta} + \text{z}\{\textbf{s}_j; \delta(\textbf{s}_j)\} \theta\left(\textbf{s}_j\right)
    \end{aligned}
    \end{equation*}
    where $r$ is the dispersion/shape parameter and $p_i(\textbf{s}_j)$ is the probability parameter.
\end{itemize}

\subsubsection{Polya-Gamma latent variables}
Although each choice of likelihood has a different mathematical form, we can obtain general results for parameter updates using the results of \citeauthor{polson2013bayesian} (2013). Specifically, when observation-level $\text{P\'olya-Gamma}$ (PG) random variables are introduced, the binomial and negative binomial likelihoods can be rewritten to allow for conjugacy, depending on the selected prior distributions for the other model parameters.  

For notational convenience, we define $\psi_i(\textbf{s}_j)$ as the linear log-odds as
\begin{equation*}
    \psi_i(\textbf{s}_j) = \text{ln}\left\{\frac{p_i(\textbf{s}_j)}{1 - p_i(\textbf{s}_j)}\right\} = \text{O}_i(\textbf{s}_j) + \textbf{x}_i(\textbf{s}_j)^{\text{T}}\boldsymbol{\beta} + \text{z}\{\textbf{s}_j; \delta(\textbf{s}_j)\}\theta\left(\textbf{s}_j\right).
\end{equation*} We then introduce PG random variables such that: \begin{itemize}
    \item Binomial likelihood: $\omega_i(\textbf{s}_j) | \psi_i(\textbf{s}_j) \sim \text{PG}\{\tilde{n}_i(\textbf{s}_j), \psi_i(\textbf{s}_j)\}$; 
    \item Negative binomial likelihood: $\omega_i(\textbf{s}_j) | \psi_i(\textbf{s}_j)  \sim \text{PG}\{r + Y_i(\textbf{s}_j), \psi_i(\textbf{s}_j)\}$.
\end{itemize}
Note that for Gaussian likelihoods these auxiliary variables are not needed for conjugacy. Use of these auxiliary variables leads to the following general form:  
\begin{align*}
&f\left(\textbf{Y}|\boldsymbol{\psi}\right) f\left(\boldsymbol{\omega} | \boldsymbol{\psi}\right) \propto \text{exp} \left\{ -\frac{1}{2} \left(\boldsymbol{\psi} - \boldsymbol{\lambda} \right)^{\text{T}} \Omega \left(\boldsymbol{\psi} - \boldsymbol{\lambda} \right)\right\}
\end{align*} where $\Omega$ is a diagonal matrix with $\frac{1}{\sigma^2_{\epsilon}}$ on the diagonal for the Gaussian likelihood with identity link and $\boldsymbol{\omega}$ on the diagonal for the other likelihood and link function options;
\begin{align*}
    \lambda_{i}(\textbf{s}_j) &=
    \begin{cases}
    Y_i(\textbf{s}_j) & \text{Gaussian/identity} \\
    \left\{Y_i\left(\textbf{s}_j\right) - 0.50\right\}/\omega_i(\textbf{s}_j) &  \text{Binomial/logit} \\
    0.50\left\{Y_i(\textbf{s}_j) - r\right\}/\omega_i(\textbf{s}_j) & \text{Negative binomial/logit}
    \end{cases}
\end{align*} and $\boldsymbol{Y}$, $\boldsymbol{\omega}$, $\boldsymbol{\psi}$, $\boldsymbol{\lambda}$ are the vectors containing the corresponding individual-specific parameters sorted similarly to $\boldsymbol{\mu}$. 

\subsubsection{Likelihood-specific full conditional updates}
For notational simplicity, we denote the total sample size across all locations as $n^* = \sum_{j=1}^mn_j$. 

\subsubsection*{Gaussian:}
Assuming $\sigma^2_{\epsilon} \sim \text{Inverse Gamma}\left(a_{\sigma^2_{\epsilon}}, b_{\sigma^2_{\epsilon}}\right)$, the full conditional distribution is given as: 
\begin{align*}
f\left(\sigma^2_{\epsilon} | \text{rest}\right) &\propto f\left(\boldsymbol{Y} | \boldsymbol{\beta}, \boldsymbol{\eta}, \boldsymbol{\gamma}, \boldsymbol{\phi}, \sigma^2_{\epsilon}\right) f(\sigma^2_{\epsilon})\\ 
\sigma^2_{\epsilon} | \text{rest} &\sim \text{Inverse Gamma}\left( a_{\sigma^2_{\epsilon}} + \frac{n^*}{2},\ b_{\sigma^2_{\epsilon}} + \frac{1}{2}\left(\boldsymbol{Y} - \boldsymbol{\mu}\right)^\text{T} \left(\boldsymbol{Y} - \boldsymbol{\mu}\right) \right)
\end{align*} where $\boldsymbol{\mu} =  \textbf{X} \boldsymbol{\beta} + \widetilde{\textbf{Z}}\left(\boldsymbol{\delta}\right) \boldsymbol{\eta}$ since $g\left(.\right)$ is the identity link for Gaussian data. 

\subsubsection*{Binomial:}
$\omega_i\left(\textbf{s}_j\right)$ are updated by drawing from the appropriate Polya-Gamma distribution: 
$$
\omega_{i}(\textbf{s}_j) | \text{rest} \sim \text{PG}\left[\tilde{n}_i(\textbf{s}_j),\ \textbf{x}_i(\textbf{s}_j)^{\text{T}}\boldsymbol{\beta} + \text{z}\{\textbf{s}_j; \delta(\textbf{s}_j)\} \theta\left(\textbf{s}_j\right)\right]
$$

\subsubsection*{Negative Binomial:}
The joint full conditional distribution of the dispersion parameter $r$ and PG auxiliary variable is given as follows:
\begin{align*}
    f\left(r, \boldsymbol{\omega} | \text{rest}\right) &\propto f\left(\boldsymbol{\omega} | r, \textbf{Y}, \boldsymbol{\beta}, \boldsymbol{\eta}, \boldsymbol{\gamma}, \boldsymbol{\phi}\right) f\left(r | \textbf{Y}, \boldsymbol{\beta}, \boldsymbol{\eta}, \boldsymbol{\gamma}, \boldsymbol{\phi}\right).
\end{align*}
We use Monte Carlo sampling to draw from this joint distribution by first sampling $r$ and then using its value to sample $\boldsymbol{\omega}$. We assume a discrete uniform prior for $r$ such that $r \in [a_r, b_r]$. For each integer value of $r$ from $a_r$ to $b_r$, we calculate the likelihood of the observed data $\boldsymbol{Y}$ given $r = k$ as
\begin{align*}
p\left(r = k | \boldsymbol{Y}, \boldsymbol{\beta}, \boldsymbol{\eta}, \boldsymbol{\gamma}, \boldsymbol{\phi}\right) &\propto  f\left( \textbf{Y} | r = k, \boldsymbol{\beta}, \boldsymbol{\eta}, \boldsymbol{\gamma}, \boldsymbol{\phi} \right) \text{P}(r = k)\\
&\propto \prod_{j=1}^m \prod_{i=1}^{n_j} f\{Y_i(\textbf{s}_j) | r = k, \boldsymbol{\beta}, \boldsymbol{\eta}, \boldsymbol{\gamma}, \phi(\textbf{s}_j)\}.
\end{align*}  Due to the discrete nature of $r$, we can use this expression to sample from a categorical distribution over the range of possible $r$ values using these normalized weights as the selection probability.

To update $\omega_i(\textbf{s}_j)$, we then draw from the appropriate Polya-Gamma distribution conditional on $r$: 
$$
\boldsymbol{\omega}_{i}(\textbf{s}_j) | \text{rest} \stackrel{\text{ind}}{\sim} \text{PG}\left[r + Y_i(\textbf{s}_j),\ \text{O}_i\left(\textbf{s}_j\right) + \textbf{x}_i(\textbf{s}_j)^{\text{T}}\boldsymbol{\beta} + \text{z}\{\textbf{s}_j; \delta(\textbf{s}_j)\} \theta\left(\textbf{s}_j\right)\right].
$$

\subsubsection{Updates for all other parameters}
\noindent
1. $f\left(\boldsymbol{\beta} | \text{rest}\right) \propto 
f\left( \boldsymbol{Y} | \boldsymbol{\beta}, \boldsymbol{\eta}, \boldsymbol{\gamma}, \boldsymbol{\phi}, \boldsymbol{\zeta} \right) f\left(\boldsymbol{\omega} | \boldsymbol{\beta}, \boldsymbol{\eta}, \boldsymbol{\gamma}, \boldsymbol{\phi}, \boldsymbol{\zeta}\right) f\left(\boldsymbol{\beta}\right)$
\begin{align*}
    &\Rightarrow \boldsymbol{\beta} | \text{rest} \sim \text{MVN}\left( \boldsymbol{\mu}_{\beta}, \Sigma_{\beta} \right)\\
    &\text{ where } \Sigma_{\beta} = \left(\textbf{X}^{\text{T}} \Omega \textbf{X} + \frac{1}{\sigma^2_{\beta}} \boldsymbol{I}_{p_x} \right)^{-1}\\
    &\text{ and }\boldsymbol{\mu}_{\beta} = \Sigma_{\beta} \textbf{X}^{\text{T}} \Omega 
    \left[ \boldsymbol{\lambda} - \textbf{O} - \widetilde{\textbf{Z}}\left(\boldsymbol{\delta}\right) \boldsymbol{\eta} \right];
\end{align*} and $\boldsymbol{\zeta}$ is a vector of likelihood-specific parameters (i.e., $\sigma^2_{\epsilon}$, $r$).\\

\noindent 2. $f\left(\boldsymbol{\eta} |  \text{rest}\right) \propto f\left( \boldsymbol{Y} | \boldsymbol{\beta}, \boldsymbol{\eta},  \boldsymbol{\gamma}, \boldsymbol{\phi}, \boldsymbol{\zeta} \right) f\left(\boldsymbol{\omega} | \boldsymbol{\beta}, \boldsymbol{\eta}, \boldsymbol{\gamma}, \boldsymbol{\phi}, \boldsymbol{\zeta}\right) f\left(\boldsymbol{\eta}\right)$
\begin{align*}
&\Rightarrow \boldsymbol{\eta} | \text{rest} \sim \text{MVN}\left(\boldsymbol{\mu}_{\eta}, \Sigma_{\eta}  \right)\\
&\text{ where } \Sigma_{\eta} = \left\{ \widetilde{\textbf{Z}}\left(\boldsymbol{\delta}\right)^{\text{T}} \Omega \widetilde{\textbf{Z}}\left(\boldsymbol{\delta}\right) + \frac{1}{\sigma^2_{\eta}}\boldsymbol{I}_{p_q} \right\}^{-1}\\
&\text{ and } \boldsymbol{\mu}_{\eta} = \Sigma_{\eta}\widetilde{\textbf{Z}}\left(\boldsymbol{\delta}\right)^{\text{T}} \Omega \left(\boldsymbol{\lambda} - \textbf{O} - \textbf{X}\boldsymbol{\beta}\right).\\
\end{align*}

\noindent 3. $f\left(\gamma_k| \text{rest}\right) \propto  f\left(\boldsymbol{Y} | \boldsymbol{\beta}, \boldsymbol{\eta}, \boldsymbol{\gamma},  \boldsymbol{\phi}, \boldsymbol{\zeta} \right) f\left(\boldsymbol{\omega} | \boldsymbol{\beta}, \boldsymbol{\eta}, \boldsymbol{\gamma}, \boldsymbol{\phi}, \boldsymbol{\zeta}\right) f\left(\gamma_k | \tau_{\phi}^2\right)$
\begin{align*}
&\Rightarrow f\left(\gamma_k | \text{rest}\right) \propto\\ &\exp \left\{  -\frac{1}{2}\left( \boldsymbol{\lambda} - \textbf{O} - \textbf{X}\boldsymbol{\beta} -  \widetilde{\textbf{Z}}\left(\boldsymbol{\delta}\right) \boldsymbol{\eta}\right)^{\text{T}} \Omega \left(\boldsymbol{\lambda} - \textbf{O} - \textbf{X}\boldsymbol{\beta} -  \widetilde{\textbf{Z}}\left(\boldsymbol{\delta}\right) \boldsymbol{\eta}\right) -\frac{p_w}{2\left(1 - \tau_{\phi}^2\right)}\gamma_k^2\right\}.
\end{align*} Note that when any entry of $\boldsymbol{\gamma}$ changes, $\boldsymbol{\delta}$ changes as well, per the specification in (\ref{eq:spatial_radii}), which changes the exposure value $\text{z}\{\textbf{s}_j; \delta(\textbf{s}_j)\}$. Hence, terms involving $\widetilde{\textbf{Z}}\left(\boldsymbol{\delta}\right)$ must be retained with respect to $\gamma_k$ in the above expression. Since this expression does not readily lead to a conjugate full conditional distribution, sampling for each individual $\gamma_k$ ($k = 1,\hdots,p_w$) parameter is achieved using a Metropolis sampling step. Specifically, a new value is proposed from a Gaussian distribution centered at the current $\gamma_k$ value with a variance tuned to optimize sampling efficiency, and each $\gamma_k$ is updated individually.\\

\noindent 4. $f\left\{\phi(\textbf{s}_j) |  \text{rest}\right\} \propto  f\left( \boldsymbol{Y} | \boldsymbol{\beta}, \boldsymbol{\eta}, \boldsymbol{\gamma},  \boldsymbol{\phi}, \boldsymbol{\zeta} \right) f\left(\boldsymbol{\omega} | \boldsymbol{\beta}, \boldsymbol{\eta}, \boldsymbol{\gamma}, \boldsymbol{\phi}, \boldsymbol{\zeta}\right) f\left(\boldsymbol{\phi} | \tau^2_{\phi}, \rho_{\phi}\right)$
\begin{align*}
&\Rightarrow f\left(\phi(\textbf{s}_j) | \text{rest}\right) \propto\\ &\exp \left\{-\frac{1}{2} \left(\boldsymbol{\lambda} - \textbf{X}\boldsymbol{\beta} -  \widetilde{\textbf{Z}}\left(\boldsymbol{\delta}\right) \boldsymbol{\eta}\right)^{\text{T}} \Omega \left(\boldsymbol{\lambda} - \textbf{X}\boldsymbol{\beta} -  \widetilde{\textbf{Z}}\left(\boldsymbol{\delta}\right) \boldsymbol{\eta}\right)\right\} \exp \left\{-\frac{1}{2\tau^2_{\phi}}\boldsymbol{\phi}^\text{T} \Sigma(\rho_{\phi})^{-1}\boldsymbol{\phi}\right\}
\end{align*} Note that when $\phi(\textbf{s}_j)$ changes, $\boldsymbol{\delta}$ changes as well, per the specification in (\ref{eq:spatial_radii}), which changes the exposure value $\text{z}\{\textbf{s}_j; \delta(\textbf{s}_j)\}$. Hence, terms involving $\widetilde{\textbf{Z}}\left(\boldsymbol{\delta}\right)$ must be retained with respect to $\phi\left(\textbf{s}_j\right)$ in the above expression.  Since this expression does not readily lead to a conjugate full conditional distribution, sampling for each $\phi\left(\textbf{s}_j\right)$ is achieved using a Metropolis sampling step. Specifically, a new value is proposed from a Gaussian distribution centered at the current $\phi\left(\textbf{s}_j\right)$ value with a variance tuned to optimize sampling efficiency, and each $\phi\left(\textbf{s}_j\right)$ is updated individually. 

Additionally, recall that $\boldsymbol{\phi} | \tau^2_{\phi}, \rho_{\phi} \sim \text{MVN}\{\textbf{0}_m, \tau^2_{\phi}\Sigma(\rho_{\phi})\}$, and with the computational approximation $\boldsymbol{\phi}^* |\tau^2_{\phi}, \rho_{\phi} \sim \text{MVN}\{\textbf{0}_K, \tau^2_{\phi}\Sigma^*(\rho_{\phi})\}$. Therefore, the general form of the full conditional distribution remains the same for $\boldsymbol{\phi}$ or $\boldsymbol{\phi}^*$.\\

\noindent 5. $\rho_{\phi} \in \left(0, \infty\right)$; we apply the transformation $\pi = \ln \left( \rho_{\phi}\right) \in \Re$ and perform the update with respect to $\pi$: 
\begin{align*}
&f\left(\pi |  \text{rest}\right) \propto f\left(\boldsymbol{\phi} | \tau^2_{\phi}, \rho_{\phi}\right) f(\pi)\\
&\Rightarrow \pi | \text{rest} \propto  \left| \Sigma(\rho_{\phi})^{-1}\right|^{1/2} \exp \left\{-\frac{1}{2\tau^2_{\phi}}\boldsymbol{\phi}^{\text{T}} \Sigma(\rho_{\phi})^{-1}\boldsymbol{\phi}\right\} \exp\left\{\pi\right\}^{a_{\rho_{\phi}}} \exp \left[-b_{\rho_{\phi}} \exp\left\{\pi\right\}\right].
\end{align*}
Since this expression does not readily lead to a conjugate full conditional distribution, sampling for $\pi$ is achieved using a Metropolis sampling step. A new value is proposed from a Gaussian distribution centered at the current $\pi$ value with a variance tuned to optimize sampling efficiency.  

Note that with the computational approximation (Section \ref{sec:computational_approximation}), the vector $\widetilde{\boldsymbol{\phi}}$ depends on $\rho_{\phi}$ (i.e., $\tilde{\phi}(\textbf{s}_j) = \textbf{c}(\textbf{s}_j; \rho_{\phi})^{\text{T}} \Sigma^*(\rho_{\phi})^{-1} \boldsymbol{\phi}^*$). As a result, when $\rho_{\phi}$ changes,  $\boldsymbol{\delta}$ changes, and thus $\widetilde{\textbf{Z}}\left(\boldsymbol{\delta}\right)$ changes. Therefore, updates of $\rho_{\phi}$ when using the computational approximation follow a slightly different form: 
\begin{align*}
&f\left(\pi | \text{rest}\right) \propto f\left(\boldsymbol{Y} | \boldsymbol{\beta}, \boldsymbol{\eta}, \boldsymbol{\gamma},  \boldsymbol{\phi}, \rho_{\phi}, \boldsymbol{\zeta} \right) f\left(\boldsymbol{\omega} | \boldsymbol{\beta}, \boldsymbol{\eta}, \boldsymbol{\gamma}, \boldsymbol{\phi}, \rho_{\phi}, \boldsymbol{\zeta}\right) f\left(\boldsymbol{\phi}^* | \tau^2_{\phi}, \rho_{\phi}\right)  f\left(\pi\right)
\end{align*} with a Metropolis sampling step similarly used.

\noindent 6. $\tau_{\phi} \in \left(0, 1\right)$; we apply the transformation $\alpha = \ln \left(\frac{\tau_{\phi}}{1 - \tau_{\phi}}\right) \in \Re$ and perform the update with respect to $\alpha$: 
\begin{align*}
&f\left(\alpha |  \text{rest}\right) \propto f\left(\boldsymbol{\gamma} | \tau^2_{\phi}\right) f\left(\boldsymbol{\phi} | \tau^2_{\phi}, \rho_{\phi}\right) f(\alpha)\\
&\Rightarrow \alpha | \text{rest} \propto \left(\frac{1}{1 - \tau^2_{\phi}}\right)^{\frac{p_w}{2}}\exp\left\{-\frac{p_w}{2\left(1- \tau^2_{\phi}\right)}\sum_{j=1}^{p_w} \gamma_j^2\right\}\times\\ 
&\left(\frac{1}{\tau^2_{\phi}}\right)^{\frac{m}{2}}  \exp \left\{-\frac{1}{2\tau^2_{\phi}}\boldsymbol{\phi}^{\text{T}} \Sigma(\rho_{\phi})^{-1}\boldsymbol{\phi}\right\}\frac{\exp\left\{-\alpha\right\}}{\left(1 + \exp\left\{-\alpha\right\}\right)^2}.
\end{align*}
Since this expression does not readily lead to a conjugate full conditional distribution, sampling for $\alpha$ is achieved using a Metropolis sampling step. A new value is proposed from a Gaussian distribution centered at the current $\alpha$ value with a variance tuned to optimize sampling efficiency.  When using the computational approximation, replace $m$ with $K$ in the above equation.
\clearpage

\section{Simulation study details and results}

\begin{table}[ht]
\centering
\caption{Parameter values used to generate datasets in the simulation study.  SVBR$\left[\delta\left(\textbf{s}_j\right), \theta\right]$ and SVBR$\left[\delta\left(\textbf{s}_j\right), \theta\left(\textbf{s}_j\right)\right]$ posterior distributions are obtained from fitting the models to the Madagascar (Toliara Province) antenatal care data using the counts exposure definition. For parameter values that are defined using random samples from their posterior distribution and change with each dataset, the posterior median and $95\%$ highest posterior density interval for the distribution are presented. Otherwise, the specific value is shown.}
\label{table:sim_param_values}
\renewcommand{\arraystretch}{1.3}
\resizebox{1\textwidth}{!}{
\begin{tabular}{lllr}
\hline
\textbf{Parameter} & \textbf{Distribution/Data Source} & \textbf{Summary} &\textbf{Value}   \\
\hline
\multicolumn{3}{l}{\textbf{All Settings (1-5)}} \\
$(a, b)$ & (minimum, maximum) radii (km) & Fixed & $(0, 20)$\\
$\beta_0$ & SVBR$\left[\delta\left(\textbf{s}_j\right), \theta\right]$ $\beta_{\text{intercept}}$ posterior & Median & $-0.621$\\
$\beta_1$ & SVBR$\left[\delta\left(\textbf{s}_j\right), \theta\right]$ $\beta_{\text{employed}}$ posterior & Median & $0.413$\\
%\hline
$\gamma_0$ & SVBR$\left[\delta\left(\textbf{s}_j\right), \theta\right]$ $\gamma_0$ posterior & Random draw & $-0.005 \hspace{0.1cm} (-0.269, 0.262)$ \\
$\text{z}\{\textbf{s}_j, \delta\left(\textbf{s}_j\right)\}$ & Count of facilities within $\delta(\textbf{s}_j)$ & -&  Depends on $\textbf{s}_j$ and $\delta(\textbf{s}_j)$ \\
\hline 
\multicolumn{3}{l}{\textbf{Varying Radii Settings (3-4)}} \\
$\Sigma(\rho_{\phi})$ & $\exp\left\{-\rho_{\phi} || \textbf{s}_j - \textbf{s}_{j'}||\right\}$ & - & Depends on $|| \textbf{s}_j - \textbf{s}_{j'}||$ \\
$\text{w}_1(\textbf{s}_j)$ & Observed GHFI value at $\textbf{s}_j$ & - & Depends on $\textbf{s}_j$ \\
$\tau_{\phi}$ & SVBR$\left[\delta\left(\textbf{s}_j\right), \theta\right]$ $\tau_{\phi}$ posterior & Median & $0.935$\\
$\rho_{\phi}$ & SVBR$\left[\delta\left(\textbf{s}_j\right), \theta\right]$ $\rho_{\phi}$ posterior & Median & $8.278$\\
$\gamma_1$ & SVBR$\left[\delta\left(\textbf{s}_j\right), \theta\right]$ $\gamma_{\text{GHFI}}$ posterior & Median &  $-0.232$\\
\hline
\multicolumn{3}{l}{\textbf{Single Exposure Effect Settings (1-3)}} \\
$\eta_0$  & SVBR$\left[\delta\left(\textbf{s}_j\right), \theta\right]$ $\eta_0$ posterior & Random draw & $0.619 \hspace{0.1cm} (0.465, 0.788)$\\
\hline
\multicolumn{3}{l}{\textbf{Varying Exposure Effect Setting (4)}} \\
$\text{q}_1(\textbf{s}_j)$ & Observed GHFI value at $\textbf{s}_j$ & - & Depends on $\textbf{s}_j$ \\
$\eta_0$ & SVBR$\left[\delta\left(\textbf{s}_j\right), \theta\left(\textbf{s}_j\right)\right]$ $\eta_0$ posterior & Random draw & $0.666 \hspace{0.1cm} (0.477, 0.906)$\\
$\eta_1$ & SVBR$\left[\delta\left(\textbf{s}_j\right), \theta\left(\textbf{s}_j\right)\right]$ $\eta_{\text{GHFI}}$ posterior & Median & $-0.153$\\
\hline
\multicolumn{3}{l}{\textbf{Misspecified Setting (5)}} \\
$\text{w}_1(\textbf{s}_j)$ & Observed GHFI value at $\textbf{s}_j$ & - & Depends on $\textbf{s}_j$ \\
$\text{w}_2(\textbf{s}_j)$ & Observed elevation value at $\textbf{s}_j$ & - & Depends on $\textbf{s}_j$ \\
$\tau_{\phi}$ & SVBR$\left[\delta\left(\textbf{s}_j\right), \theta\right]$ $\rho_{\phi}$ posterior & Median & $0.935$\\
$\rho_{\phi}$ & SVBR$\left[\delta\left(\textbf{s}_j\right), \theta\right]$ $\rho_{\phi}$ posterior & Median & $8.278$\\
$\gamma_1$ & SVBR$\left[\delta\left(\textbf{s}_j\right), \theta\right]$ $\gamma_{\text{GHFI}}$ posterior & Median &  $-0.232$\\
$\gamma_2$ & SVBR$\left[\delta\left(\textbf{s}_j\right), \theta\right]$ $\gamma_{\text{elev}}$ posterior & Median & $-0.099$\\
$\theta(\textbf{s}_j)$ & $\pi_0 + \pi_1\text{logit}\left\{\frac{\delta\left(\textbf{s}_j\right)}{b-a}\right\}$ & $\pi_0$, $\pi_1$: Coefficients of the linear model  & $\pi_0 = 0.653$, \\
& &   $\hat{\theta}(\textbf{s}_j) \sim \text{logit}\left\{\hat{\delta}(\textbf{s}_j)\right\}$, where $\hat{\theta}(\textbf{s}_j)$ and $ \hat{\delta}(\textbf{s}_j)$ are posterior  & $\pi_1 = -0.096$ \\
& &  median estimates of exposure effects and radii & \\
& & from SVBR$\left[\delta\left(\textbf{s}_j\right), \theta\left(\textbf{s}_j\right)\right]$ \\
$\zeta(\textbf{s}_j)$ & $\mathbf{B}_{\text{tps}} \boldsymbol{\alpha}$, where $\boldsymbol{\alpha} \sim \text{MVN}\left(\boldsymbol{0}_{30}, 
\frac{\tau_{\phi}^2}{30}\mathbf{I}_{30}\right)$ & $\mathbf{B}_{\text{tps}}$: Thin-plate spline design matrix evaluated at   & Depends on $\textbf{s}_j$\\
& & observed $\textbf{s}_j$ & \\
\hline
\end{tabular}
}
\end{table}
\clearpage

\begin{table}[h!]
\centering
\caption{\textbf{Bias} of posterior median estimates in the simulation study with \textbf{Bernoulli outcome}. For each dataset, bias is computed for each parameter and averaged across spatial locations when applicable; these values are then averaged across all $500$ simulated datasets. Standard errors are given in parentheses. Bolded values indicate the bias closest to zero in each row.}\label{table:sim_bias}
\centering
\begin{tabular}[t]{lccc}
\toprule
\multicolumn{1}{c}{ } & \multicolumn{3}{c}{Model Type} \\
\cmidrule(l{3pt}r{3pt}){2-4}
Parameter & $\text{SVBR}[\delta, \theta]$ & $\text{SVBR}[\delta(\textbf{s}_j), \theta]$ & $\text{SVBR}[\delta(\textbf{s}_j), \theta(\textbf{s}_j)]$\\
\midrule
\addlinespace[0.3em]
\multicolumn{4}{l}{\textbf{No Effect}}\\
\hspace{1em}$\text{z}\left(\textbf{s}_j; \delta\right) \theta$ & -0.01 (0.00) & \textbf{-0.00} (0.00) & -0.01 (0.00)\\
\hspace{1em}$\beta_1$ & \textbf{0.01} (0.01) & 0.01 (0.01) & 0.01 (0.01)\\
\addlinespace[0.3em]
\multicolumn{4}{l}{\textbf{Single Radius \& Effect}}\\
\hspace{1em}$\delta$ & \textbf{0.14} (0.04) & 0.16 (0.04) & 0.16 (0.04)\\
\hspace{1em}$\theta$ & \textbf{0.01} (0.00) & 0.02 (0.00) & 0.03 (0.01)\\
\hspace{1em}$\text{z}\left(\textbf{s}_j; \delta\right) \theta$ & \textbf{0.03} (0.01) & 0.04 (0.01) & 0.11 (0.01)\\
\hspace{1em}$\beta_1$ & \textbf{0.01} (0.01) & 0.02 (0.01) & 0.02 (0.01)\\
\addlinespace[0.3em]
\multicolumn{4}{l}{\textbf{Varying Radii, Single Effect}}\\
\hspace{1em}$\delta(\textbf{s}_j)$ & -1.08 (0.23) & 0.14 (0.09) & \textbf{-0.14} (0.09)\\
\hspace{1em}$\theta$ & 0.12 (0.10) & -0.02 (0.01) & \textbf{0.01} (0.01)\\
\hspace{1em}$\text{z}\left\{\textbf{s}_j; \delta(\textbf{s}_j)\right\} \theta$ & -0.41 (0.01) & -0.09 (0.01) & \textbf{-0.05} (0.01)\\
\hspace{1em}$\beta_1$ & -0.03 (0.01) & \textbf{0.01} (0.01) & 0.01 (0.01)\\
\hspace{1em}$\gamma_1$ & -- & 0.06 (0.01) & \textbf{0.05} (0.01)\\
\addlinespace[0.3em]
\multicolumn{4}{l}{\textbf{Varying Radii \& Effect}}\\
\hspace{1em}$\delta(\textbf{s}_j)$ & -1.53 (0.27) & 0.43 (0.10) & \textbf{-0.05} (0.09)\\
\hspace{1em}$\theta(\textbf{s}_j)$ & 0.36 (0.17) & -0.04 (0.01) & \textbf{0.01} (0.01)\\
\hspace{1em}$\text{z}\left\{\textbf{s}_j; \delta(\textbf{s}_j)\right\} \theta\left(\textbf{s}_j\right)$ & -0.49 (0.01) & -0.08 (0.01) & \textbf{-0.06} (0.01)\\
\hspace{1em}$\beta_1$ & -0.03 (0.01) & -0.02 (0.01) & \textbf{-0.01} (0.01)\\
\hspace{1em}$\gamma_1$ & -- & -0.21 (0.01) & \textbf{-0.04} (0.01)\\
\hspace{1em}$\eta_1$ & -- & -- & \textbf{0.06} (0.01)\\
\addlinespace[0.3em]
\multicolumn{4}{l}{\textbf{Misspecified}}\\
\hspace{1em}$\delta(\textbf{s}_j)$ & -0.09 (0.11) & \textbf{0.01} (0.06) & -0.15 (0.07)\\
\hspace{1em}$\theta(\textbf{s}_j)$ & -0.09 (0.01) & -0.02 (0.01) & \textbf{0.00} (0.01)\\
\hspace{1em}$\text{z}\left\{\textbf{s}_j; \delta(\textbf{s}_j)\right\} \theta\left(\textbf{s}_j\right)$ & -0.16 (0.01) & -0.03 (0.01) & \textbf{0.02} (0.01)\\
\hspace{1em}$\beta_1$ & \textbf{0.01} (0.01) & 0.03 (0.01) & 0.03 (0.01)\\
\hspace{1em}$\gamma_1$ & -- & 0.23 (0.01) & \textbf{0.18} (0.01)\\
\bottomrule
\end{tabular}
\end{table}

\begin{table}
\centering
\caption{\textbf{Empirical coverage (EC)} of the 95\% highest posterior density credible intervals in the simulation study with \textbf{Bernoulli outcome}. For each dataset, EC is computed for each parameter and averaged across spatial locations when applicable; these values are then averaged across all $500$ simulated datasets. Standard errors are given in parentheses. Bolded values indicate the EC closest to $0.95$ in each row.}\label{table:sim_ec}
\centering
\begin{tabular}[t]{lccc}
\toprule
\multicolumn{1}{c}{ } & \multicolumn{3}{c}{Model Type} \\
\cmidrule(l{3pt}r{3pt}){2-4}
Parameter & $\text{SVBR}[\delta, \theta]$ & $\text{SVBR}[\delta(\textbf{s}_j), \theta]$ & $\text{SVBR}[\delta(\textbf{s}_j), \theta(\textbf{s}_j)]$\\
\midrule
\addlinespace[0.3em]
\multicolumn{4}{l}{\textbf{No Effect}}\\
\hspace{1em}$\text{z}\left(\textbf{s}_j; \delta\right) \theta$ & \textbf{1.00} (0.00) & \textbf{1.00} (0.00) & \textbf{1.00} (0.00)\\
\hspace{1em}$\beta_1$ & \textbf{0.96} (0.01) & \textbf{0.96} (0.01) & \textbf{0.96} (0.01)\\
\addlinespace[0.3em]
\multicolumn{4}{l}{\textbf{Single Radius \& Effect}}\\
\hspace{1em}$\delta$ & \textbf{0.94} (0.01) & 0.99 (0.00) & 0.98 (0.01)\\
\hspace{1em}$\theta$ & 0.94 (0.01) & \textbf{0.95} (0.01) & 0.96 (0.01)\\
\hspace{1em}$\text{z}\left(\textbf{s}_j; \delta\right) \theta$ & 0.97 (0.01) & 1.00 (0.00) & 1.00 (0.00)\\
\hspace{1em}$\beta_1$ & \textbf{0.95} (0.01) & 0.94 (0.01) & 0.94 (0.01)\\
\addlinespace[0.3em]
\multicolumn{4}{l}{\textbf{Varying Radii, Single Effect}}\\
\hspace{1em}$\delta(\textbf{s}_j)$ & 0.28 (0.02) & 0.85 (0.02) & \textbf{0.87} (0.01)\\
\hspace{1em}$\theta$ & 0.60 (0.02) & 0.87 (0.02) & \textbf{0.92} (0.01)\\
\hspace{1em}$\text{z}\left\{\textbf{s}_j; \delta(\textbf{s}_j)\right\} \theta$ & 0.60 (0.02) & \textbf{0.96} (0.01) & 0.98 (0.01)\\
\hspace{1em}$\beta_1$ & \textbf{0.94} (0.01) & \textbf{0.94} (0.01) & \textbf{0.94} (0.01)\\
\hspace{1em}$\gamma_1$ & -- & 0.87 (0.02) & \textbf{0.94} (0.01)\\
\addlinespace[0.3em]
\multicolumn{4}{l}{\textbf{Varying Radii \& Effect}}\\
\hspace{1em}$\delta(\textbf{s}_j)$ & 0.35 (0.02) & 0.82 (0.02) & \textbf{0.87} (0.02)\\
\hspace{1em}$\theta(\textbf{s}_j)$ & 0.48 (0.02) & 0.82 (0.02) & \textbf{0.90} (0.01)\\
\hspace{1em}$\text{z}\left\{\textbf{s}_j; \delta(\textbf{s}_j)\right\} \theta\left(\textbf{s}_j\right)$ & 0.41 (0.02) & 0.97 (0.01) & 0.99 (0.00)\\
\hspace{1em}$\beta_1$ & 0.92 (0.01) & \textbf{0.94} (0.01) & \textbf{0.94} (0.01)\\
\hspace{1em}$\gamma_1$ & -- & 0.61 (0.02) & \textbf{0.91 }(0.01)\\
\hspace{1em}$\eta_1$ & -- & -- & \textbf{0.93} (0.01)\\
\addlinespace[0.3em]
\multicolumn{4}{l}{\textbf{Misspecified}}\\
\hspace{1em}$\delta(\textbf{s}_j)$ & 0.32 (0.02) & \textbf{0.95} (0.01) & 0.93 (0.01)\\
\hspace{1em}$\theta(\textbf{s}_j)$ & 0.76 (0.02) & 0.89 (0.01) & \textbf{0.93} (0.01)\\
\hspace{1em}$\text{z}\left\{\textbf{s}_j; \delta(\textbf{s}_j)\right\} \theta\left(\textbf{s}_j\right)$ & \textbf{0.93} (0.01) & 1.00 (0.00) & 1.00 (0.00)\\
\hspace{1em}$\beta_1$ & \textbf{0.95} (0.01) & 0.96 (0.01) & \textbf{0.95} (0.01)\\
\hspace{1em}$\gamma_1$ & -- & 0.81 (0.02) & \textbf{0.84} (0.02)\\
\bottomrule
\end{tabular}
\end{table}
\clearpage 

\begin{table}
\centering
\caption{\textbf{Length} of the 95\% highest posterior density credible intervals in the simulation study with \textbf{Bernoulli outcome}. For each dataset, length is computed for each parameter and averaged across spatial locations when applicable; these values are then averaged across all $500$ simulated datasets. Standard errors are given in parentheses.}\label{table:sim_hdi_length}
\centering
\begin{tabular}[t]{lccc}
\toprule
\multicolumn{1}{c}{ } & \multicolumn{3}{c}{Model Type} \\
\cmidrule(l{3pt}r{3pt}){2-4}
Parameter & $\text{SVBR}[\delta, \theta]$ & $\text{SVBR}[\delta(\textbf{s}_j), \theta]$ & $\text{SVBR}[\delta(\textbf{s}_j), \theta(\textbf{s}_j)]$\\
\midrule
\addlinespace[0.3em]
\multicolumn{4}{l}{\textbf{No Effect}}\\
\hspace{1em}$\text{z}\left(\textbf{s}_j; \delta\right) \theta$ & 0.14 (0.00) & 0.38 (0.00) & 0.27 (0.01)\\
\hspace{1em}$\beta_1$ & 0.99 (0.00) & 0.99 (0.00) & 0.99 (0.00)\\
\addlinespace[0.3em]
\multicolumn{4}{l}{\textbf{Single Radius \& Effect}}\\
\hspace{1em}$\delta$ & 2.48 (0.06) & 5.39 (0.10) & 5.59 (0.09)\\
\hspace{1em}$\theta$ & 0.41 (0.00) & 0.44 (0.00) & 0.51 (0.00)\\
\hspace{1em}$\text{z}\left(\textbf{s}_j; \delta\right) \theta$ & 0.44 (0.00) & 0.61 (0.01) & 0.70 (0.01)\\
\hspace{1em}$\beta_1$ & 1.09 (0.00) & 1.10 (0.00) & 1.11 (0.00)\\
\addlinespace[0.3em]
\multicolumn{4}{l}{\textbf{Varying Radii, Single Effect}}\\
\hspace{1em}$\delta(\textbf{s}_j)$ & 6.77 (0.23) & 9.88 (0.12) & 10.02 (0.12)\\
\hspace{1em}$\theta$ & 1.71 (0.34) & 0.55 (0.01) & 0.90 (0.12)\\
\hspace{1em}$\text{z}\left\{\textbf{s}_j; \delta(\textbf{s}_j)\right\} \theta$ & 0.38 (0.01) & 0.88 (0.01) & 0.95 (0.01)\\
\hspace{1em}$\beta_1$ & 1.04 (0.00) & 1.09 (0.00) & 1.09 (0.00)\\
\hspace{1em}$\gamma_1$ & -- & 0.70 (0.01) & 0.83 (0.01)\\
\addlinespace[0.3em]
\multicolumn{4}{l}{\textbf{Varying Radii \& Effect}}\\
\hspace{1em}$\delta(\textbf{s}_j)$ & 7.47 (0.25) & 9.24 (0.12) & 9.74 (0.12)\\
\hspace{1em}$\theta(\textbf{s}_j)$ & 2.98 (0.50) & 0.59 (0.01) & 1.00 (0.16)\\
\hspace{1em}$\text{z}\left\{\textbf{s}_j; \delta(\textbf{s}_j)\right\} \theta\left(\textbf{s}_j\right)$ & 0.30 (0.01) & 0.89 (0.01) & 0.98 (0.02)\\
\hspace{1em}$\beta_1$ & 1.02 (0.00) & 1.08 (0.00) & 1.08 (0.00)\\
\hspace{1em}$\gamma_1$ & -- & 0.69 (0.02) & 0.92 (0.02)\\
\hspace{1em}$\eta_1$ & -- & -- & 1.09 (0.28)\\
\addlinespace[0.3em]
\multicolumn{4}{l}{\textbf{Misspecified}}\\
\hspace{1em}$\delta(\textbf{s}_j)$ & 4.23 (0.13) & 8.29 (0.10) & 7.94 (0.10)\\
\hspace{1em}$\theta(\textbf{s}_j)$ & 0.50 (0.01) & 0.51 (0.01) & 0.60 (0.01)\\
\hspace{1em}$\text{z}\left\{\textbf{s}_j; \delta(\textbf{s}_j)\right\} \theta\left(\textbf{s}_j\right)$ & 0.49 (0.01) & 0.82 (0.01) & 0.84 (0.01)\\
\hspace{1em}$\beta_1$ & 1.07 (0.00) & 1.10 (0.00) & 1.10 (0.00)\\
\hspace{1em}$\gamma_1$ & -- & 0.78 (0.01) & 0.68 (0.01)\\
\bottomrule
\end{tabular}
\end{table}
\clearpage

\begin{table}
\centering
\caption{Proportion of the $500$ simulated datasets favored by each competing method in the \textbf{Bernoulli outcome} simulation, with respect to Watanabe-Akaike information criteria (WAIC), the Pareto-smoothed importance sampling leave-one-out (PSIS-LOO) cross validation, and the logarithm of the pseudo marginal likelihood (LPML), across all simulation settings. (Note, proportions may not sum to one due to rounding).}\label{table:sim_lowest_waic}
\begin{tabular}[t]{lccc}
\toprule
Simulation Setting & SVBR$\left[\delta, \theta\right]$ & SVBR$\left[\delta\left(\textbf{s}_j\right), \theta\right]$ & SVBR$\left[\delta\left(\textbf{s}_j\right), \theta\left(\textbf{s}_j\right)\right]$\\
\midrule
\textbf{No Effect}               &      &      &     \\
\hspace{1em} WAIC                & 0.32 & 0.49 & 0.19\\
\hspace{1em} LPML           &   0.30   &   0.55   &   0.14  \\
\hspace{1em} PSIS-LOO                &    0.30  &   0.55   &  0.15   \\
\textbf{Single Radius \& Effect}          &      &      &     \\
\hspace{1em} WAIC                & 0.66 & 0.23 & 0.12\\
\hspace{1em} LPML            &    0.66  &   0.22   &  0.12   \\
\hspace{1em} PSIS-LOO                &  0.66  &   0.22   &  0.12    \\
\textbf{Varying Radii, Single Effect}   &      &      &     \\
\hspace{1em} WAIC                & 0.05 & 0.57 & 0.37\\
\hspace{1em} LPML            &  0.06    &  0.59    &  0.35   \\
\hspace{1em} PSIS-LOO                & 0.06    &  0.59    &  0.35   \\
\textbf{Varying Radii \& Effect}          &      &      &     \\
\hspace{1em} WAIC                & 0.02 & 0.43 & 0.55\\
\hspace{1em} LPML            &    0.02  &  0.46    &  0.52   \\
\hspace{1em} PSIS-LOO                &   0.02   &  0.44    & 0.53    \\
\textbf{Misspecified} &      &      &     \\
\hspace{1em} WAIC                & 0.17 & 0.50 & 0.33\\
\hspace{1em} LPML            &   0.18   &   0.50   & 0.32    \\
\hspace{1em} PSIS-LOO                &    0.18  &  0.50    &  0.32   \\
\bottomrule
\end{tabular}
\end{table}
\clearpage

\begin{table}[h!]
\centering
\caption{\textbf{Mean absolute error (MAE)} of posterior median estimates in the simulation study with \textbf{Gaussian outcome}. For each dataset, MAE is computed for each parameter and averaged across spatial locations when applicable; these values are then averaged across all $500$ simulated datasets. Standard errors are given in parentheses. Bolded values indicate the smallest MAE in each row.}\label{table:sim_gauss_mae}
\centering
\begin{tabular}[t]{lccc}
\toprule
\multicolumn{1}{c}{ } & \multicolumn{3}{c}{Model Type} \\
\cmidrule(l{3pt}r{3pt}){2-4}
Parameter & $\text{SVBR}[\delta, \theta]$ & $\text{SVBR}[\delta(\textbf{s}_j), \theta]$ & $\text{SVBR}[\delta(\textbf{s}_j), \theta(\textbf{s}_j)]$\\
\midrule
\addlinespace[0.3em]
\multicolumn{4}{l}{\textbf{No Effect}}\\
\hspace{1em}$\text{z}\left(\textbf{s}_j; \delta\right) \theta$ & \textbf{0.02} (0.00) & 0.03 (0.00) & 0.03 (0.00)\\
\hspace{1em}$\beta_1$ & 0.09 (0.00) & \textbf{0.09} (0.00) & 0.09 (0.00)\\
\addlinespace[0.3em]
\multicolumn{4}{l}{\textbf{Single Radius \& Effect}}\\
\hspace{1em}$\delta$ & \textbf{0.08} (0.00) & 0.27 (0.01) & 0.30 (0.01)\\
\hspace{1em}$\theta$ & \textbf{0.01} (0.00) & 0.02 (0.00) & 0.03 (0.00)\\
\hspace{1em}$\text{z}\left(\textbf{s}_j; \delta\right) \theta$ & \textbf{0.03} (0.00) & 0.05 (0.00) & 0.06 (0.00)\\
\hspace{1em}$\beta_1$ & \textbf{0.10} (0.00) & 0.10 (0.00) & 0.10 (0.00)\\
\addlinespace[0.3em]
\multicolumn{4}{l}{\textbf{Varying Radii, Single Effect}}\\
\hspace{1em}$\delta(\textbf{s}_j)$ & 5.64 (0.08) & \textbf{2.50} (0.02) & 2.55 (0.02)\\
\hspace{1em}$\theta$ & 0.21 (0.01) & \textbf{0.03} (0.00) & 0.05 (0.00)\\
\hspace{1em}$\text{z}\left\{\textbf{s}_j; \delta(\textbf{s}_j)\right\} \theta$ & 0.56 (0.01) & \textbf{0.15} (0.00) & 0.16 (0.00)\\
\hspace{1em}$\beta_1$ & 0.15 (0.01) & 0.10 (0.00) & \textbf{0.10} (0.00)\\
\hspace{1em}$\gamma_1$ & -- & \textbf{0.10} (0.00) & 0.11 (0.00)\\
\addlinespace[0.3em]
\multicolumn{4}{l}{\textbf{Varying Radii \& Effect}}\\
\hspace{1em}$\delta(\textbf{s}_j)$ & 6.58 (0.09) & 2.86 (0.02) & \textbf{2.51} (0.02)\\
\hspace{1em}$\theta(\textbf{s}_j)$ & 0.43 (0.01) & 0.12 (0.00) & \textbf{0.04} (0.00)\\
\hspace{1em}$\text{z}\left\{\textbf{s}_j; \delta(\textbf{s}_j)\right\} \theta\left(\textbf{s}_j\right)$ & 0.69 (0.01) & 0.20 (0.00) & \textbf{0.15} (0.00)\\
\hspace{1em}$\beta_1$ & 0.16 (0.01) & 0.10 (0.00) & \textbf{0.10} (0.00)\\
\hspace{1em}$\gamma_1$ & -- & 0.25 (0.01) & \textbf{0.12} (0.00)\\
\hspace{1em}$\eta_1$ & -- & -- & \textbf{0.03} (0.00)\\
\addlinespace[0.3em]
\multicolumn{4}{l}{\textbf{Misspecified}}\\
\hspace{1em}$\delta(\textbf{s}_j)$ & 3.15 (0.05) & 1.69 (0.02) & \textbf{1.64} (0.02)\\
\hspace{1em}$\theta(\textbf{s}_j)$ & 0.11 (0.00) & \textbf{0.07} (0.00) & 0.08 (0.00)\\
\hspace{1em}$\text{z}\left\{\textbf{s}_j; \delta(\textbf{s}_j)\right\} \theta\left(\textbf{s}_j\right)$ & 0.31 (0.01) & 0.14 (0.00) & \textbf{0.14} (0.00)\\
\hspace{1em}$\beta_1$ & 0.12 (0.00) & 0.10 (0.00) & \textbf{0.10} (0.00)\\
\hspace{1em}$\gamma_1$ & -- & 0.18 (0.00) & \textbf{0.14} (0.00)\\
\bottomrule
\end{tabular}
\end{table}

\begin{table}[h!]
\centering
\caption{\textbf{Bias} of posterior median estimates in the simulation study with \textbf{Gaussian outcome}. For each dataset, bias is computed for each parameter and averaged across spatial locations when applicable; these values are then averaged across all $500$ simulated datasets. Standard errors are given in parentheses. Bolded values indicate the bias closest to zero in each row.}\label{table:sim_gauss_bias}
\centering
\begin{tabular}[t]{lccc}
\toprule
\multicolumn{1}{c}{ } & \multicolumn{3}{c}{Model Type} \\
\cmidrule(l{3pt}r{3pt}){2-4}
Parameter & $\text{SVBR}[\delta, \theta]$ & $\text{SVBR}[\delta(\textbf{s}_j), \theta]$ & $\text{SVBR}[\delta(\textbf{s}_j), \theta(\textbf{s}_j)]$\\
\midrule
\addlinespace[0.3em]
\multicolumn{4}{l}{\textbf{No Effect}}\\
\hspace{1em}$\text{z}\left(\textbf{s}_j; \delta\right) \theta$ & \textbf{0.00} (0.00) & 0.00 (0.00) & 0.00 (0.00)\\
\hspace{1em}$\beta_1$ & 0.00 (0.01) & 0.00 (0.01) & \textbf{0.00} (0.01)\\
\addlinespace[0.3em]
\multicolumn{4}{l}{\textbf{Single Radius \& Effect}}\\
\hspace{1em}$\delta$ & \textbf{0.01} (0.01) & 0.01 (0.01) & 0.01 (0.01)\\
\hspace{1em}$\theta$ & \textbf{0.00} (0.00) & -0.00 (0.00) & 0.00 (0.00)\\
\hspace{1em}$\text{z}\left(\textbf{s}_j; \delta\right) \theta$ & \textbf{0.00} (0.00) & -0.00 (0.00) & 0.00 (0.00)\\
\hspace{1em}$\beta_1$ & 0.00 (0.01) & \textbf{0.00} (0.01) & 0.00 (0.01)\\
\addlinespace[0.3em]
\multicolumn{4}{l}{\textbf{Varying Radii, Single Effect}}\\
\hspace{1em}$\delta(\textbf{s}_j)$ & 0.38 (0.22) & \textbf{0.10} (0.04) & 0.10 (0.04)\\
\hspace{1em}$\theta$ & -0.11 (0.01) & \textbf{0.00} (0.00) & 0.01 (0.00)\\
\hspace{1em}$\text{z}\left\{\textbf{s}_j; \delta(\textbf{s}_j)\right\} \theta$ & -0.20 (0.02) & 0.00 (0.00) & \textbf{0.00} (0.00)\\
\hspace{1em}$\beta_1$ & -0.01 (0.01) & \textbf{0.00} (0.01) & 0.00 (0.01)\\
\hspace{1em}$\gamma_1$ & -- & \textbf{0.02} (0.01) & 0.04 (0.01)\\
\addlinespace[0.3em]
\multicolumn{4}{l}{\textbf{Varying Radii \& Effect}}\\
\hspace{1em}$\delta(\textbf{s}_j)$ & 1.19 (0.28) & \textbf{0.09} (0.04) & 0.11 (0.03)\\
\hspace{1em}$\theta(\textbf{s}_j)$ & -0.25 (0.02) & -0.01 (0.00) & \textbf{0.01} (0.00)\\
\hspace{1em}$\text{z}\left\{\textbf{s}_j; \delta(\textbf{s}_j)\right\} \theta\left(\textbf{s}_j\right)$ & -0.42 (0.01) & -0.03 (0.00) & \textbf{0.00} (0.00)\\
\hspace{1em}$\beta_1$ & 0.01 (0.01) & 0.01 (0.01) & \textbf{0.00} (0.01)\\
\hspace{1em}$\gamma_1$ & -- & -0.24 (0.01) & \textbf{0.03} (0.01)\\
\hspace{1em}$\eta_1$ & -- & -- & \textbf{0.00} (0.00)\\
\addlinespace[0.3em]
\multicolumn{4}{l}{\textbf{Misspecified}}\\
\hspace{1em}$\delta(\textbf{s}_j)$ & -0.30 (0.11) & -0.16 (0.04) & \textbf{-0.01} (0.03)\\
\hspace{1em}$\theta(\textbf{s}_j)$ & -0.06 (0.01) & \textbf{-0.01} (0.00) & -0.03 (0.00)\\
\hspace{1em}$\text{z}\left\{\textbf{s}_j; \delta(\textbf{s}_j)\right\} \theta\left(\textbf{s}_j\right)$ & -0.08 (0.01) & \textbf{-0.01} (0.00) & -0.02 (0.00)\\
\hspace{1em}$\beta_1$ & -0.01 (0.01) & 0.00 (0.01) & \textbf{0.00} (0.01)\\
\hspace{1em}$\gamma_1$ & -- & 0.18 (0.00) & \textbf{0.14} (0.00)\\
\bottomrule
\end{tabular}
\end{table}
 
\begin{table}
\centering
\caption{\textbf{Empirical coverage} (EC) of the 95\% highest posterior density credible intervals in the simulation study with \textbf{Gaussian outcome}. For each dataset, EC is computed for each parameter and averaged across spatial locations when applicable; these values are then averaged across all $500$ simulated datasets. Standard errors are given in parentheses. Bolded values indicate the EC closest to $0.95$ in each row.}\label{table:sim_ec_gaussian}
\centering
\begin{tabular}[t]{lccc}
\toprule
\multicolumn{1}{c}{ } & \multicolumn{3}{c}{Model Type} \\
\cmidrule(l{3pt}r{3pt}){2-4}
Parameter & $\text{SVBR}[\delta, \theta]$ & $\text{SVBR}[\delta(\textbf{s}_j), \theta]$ & $\text{SVBR}[\delta(\textbf{s}_j), \theta(\textbf{s}_j)]$\\
\midrule
\addlinespace[0.3em]
\multicolumn{4}{l}{\textbf{No Effect}}\\
\hspace{1em}$\text{z}\left(\textbf{s}_j; \delta\right) \theta$ & \textbf{1.00} (0.00) & \textbf{1.00} (0.00) & \textbf{1.00} (0.00)\\
\hspace{1em}$\beta_1$ & \textbf{0.95} (0.01) & \textbf{0.95} (0.01) & \textbf{0.95} (0.01)\\
\addlinespace[0.3em]
\multicolumn{4}{l}{\textbf{Single Radius \& Effect}}\\
\hspace{1em}$\delta$ & \textbf{0.96} (0.01) & 0.99 (0.00) & 0.99 (0.01)\\
\hspace{1em}$\theta$ & \textbf{0.95} (0.01) & 0.94 (0.01) & \textbf{0.95} (0.01)\\
\hspace{1em}$\text{z}\left(\textbf{s}_j; \delta\right) \theta$ & \textbf{0.95} (0.01) & 0.97 (0.01) & 0.99 (0.00)\\
\hspace{1em}$\beta_1$ & \textbf{0.95} (0.01) & \textbf{0.95} (0.01) & \textbf{0.95} (0.01)\\
\addlinespace[0.3em]
\multicolumn{4}{l}{\textbf{Varying Radii, Single Effect}}\\
\hspace{1em}$\delta(\textbf{s}_j)$ & 0.06 (0.01) & \textbf{1.00} (0.00) & \textbf{1.00} (0.00)\\
\hspace{1em}$\theta$ & 0.33 (0.02) & 0.93 (0.01) & \textbf{0.95} (0.01)\\
\hspace{1em}$\text{z}\left\{\textbf{s}_j; \delta(\textbf{s}_j)\right\} \theta$ & 0.34 (0.02) & \textbf{1.00} (0.00) & \textbf{1.00} (0.00)\\
\hspace{1em}$\beta_1$ & 0.90 (0.01) & 0.94 (0.01) & \textbf{0.95} (0.01)\\
\hspace{1em}$\gamma_1$ & -- & 0.84 (0.02) & \textbf{0.88} (0.01)\\
\addlinespace[0.3em]
\multicolumn{4}{l}{\textbf{Varying Radii \& Effect}}\\
\hspace{1em}$\delta(\textbf{s}_j)$ & 0.07 (0.01) & \textbf{0.99} (0.00) & 1.00 (0.00)\\
\hspace{1em}$\theta(\textbf{s}_j)$ & 0.21 (0.02) & 0.50 (0.02) & \textbf{0.96} (0.01)\\
\hspace{1em}$\text{z}\left\{\textbf{s}_j; \delta(\textbf{s}_j)\right\} \theta\left(\textbf{s}_j\right)$ & 0.22 (0.02) & \textbf{1.00} (0.00) & \textbf{1.00} (0.00)\\
\hspace{1em}$\beta_1$ & 0.91 (0.01) & \textbf{0.95} (0.01) & \textbf{0.95} (0.01)\\
\hspace{1em}$\gamma_1$ & -- & 0.32 (0.02) & \textbf{0.93} (0.01)\\
\hspace{1em}$\eta_1$ & -- & -- & \textbf{0.94} (0.01)\\
\addlinespace[0.3em]
\multicolumn{4}{l}{\textbf{Misspecified}}\\
\hspace{1em}$\delta(\textbf{s}_j)$ & 0.02 (0.01) & \textbf{0.96} (0.01) & 0.97 (0.01)\\
\hspace{1em}$\theta(\textbf{s}_j)$ & 0.20 (0.02) & 0.54 (0.02) & \textbf{0.74} (0.02)\\
\hspace{1em}$\text{z}\left\{\textbf{s}_j; \delta(\textbf{s}_j)\right\} \theta\left(\textbf{s}_j\right)$ & 0.56 (0.02) & \textbf{1.00 }(0.00) & \textbf{1.00} (0.00)\\
\hspace{1em}$\beta_1$ & 0.94 (0.01) & \textbf{0.96} (0.01) & \textbf{0.96} (0.01)\\
\hspace{1em}$\gamma_1$ & -- & 0.15 (0.02) & \textbf{0.53} (0.02)\\
\bottomrule
\end{tabular}
\end{table}

\begin{table}
\centering
\caption{\textbf{Length} of the 95\% highest posterior density credible intervals in the simulation study with \textbf{Gaussian outcome}. For each dataset, length is computed for each parameter and averaged across spatial locations when applicable; these values are then averaged across all $500$ simulated datasets. Standard errors are given in parentheses.}\label{table:sim_hdi_length_gaussian}
\centering
\begin{tabular}[t]{lccc}
\toprule
\multicolumn{1}{c}{ } & \multicolumn{3}{c}{Model Type} \\
\cmidrule(l{3pt}r{3pt}){2-4}
Parameter & $\text{SVBR}[\delta, \theta]$ & $\text{SVBR}[\delta(\textbf{s}_j), \theta]$ & $\text{SVBR}[\delta(\textbf{s}_j), \theta(\textbf{s}_j)]$\\
\midrule
\addlinespace[0.3em]
\multicolumn{4}{l}{\textbf{No Effect}}\\
\hspace{1em}$\text{z}\left(\textbf{s}_j; \delta\right) \theta$ & 0.07 (0.00) & 0.19 (0.00) & 0.13 (0.00)\\
\hspace{1em}$\beta_1$ & 0.47 (0.00) & 0.47 (0.00) & 0.47 (0.00)\\
\addlinespace[0.3em]
\multicolumn{4}{l}{\textbf{Single Radius \& Effect}}\\
\hspace{1em}$\delta$ & 0.39 (0.01) & 1.57 (0.04) & 1.72 (0.04)\\
\hspace{1em}$\theta$ & 0.07 (0.00) & 0.09 (0.00) & 0.15 (0.00)\\
\hspace{1em}$\text{z}\left(\textbf{s}_j; \delta\right) \theta$ & 0.07 (0.00) & 0.09 (0.00) & 0.17 (0.00)\\
\hspace{1em}$\beta_1$ & 0.47 (0.00) & 0.48 (0.00) & 0.48 (0.00)\\
\addlinespace[0.3em]
\multicolumn{4}{l}{\textbf{Varying Radii, Single Effect}}\\
\hspace{1em}$\delta(\textbf{s}_j)$ & 2.02 (0.16) & 8.76 (0.06) & 8.94 (0.06)\\
\hspace{1em}$\theta$ & 0.22 (0.01) & 0.14 (0.00) & 0.18 (0.00)\\
\hspace{1em}$\text{z}\left\{\textbf{s}_j; \delta(\textbf{s}_j)\right\} \theta$ & 0.11 (0.00) & 0.56 (0.01) & 0.60 (0.01)\\
\hspace{1em}$\beta_1$ & 0.60 (0.00) & 0.49 (0.00) & 0.49 (0.00)\\
\hspace{1em}$\gamma_1$ & -- & 0.36 (0.00) & 0.46 (0.01)\\
\addlinespace[0.3em]
\multicolumn{4}{l}{\textbf{Varying Radii \& Effect}}\\
\hspace{1em}$\delta(\textbf{s}_j)$ & 2.69 (0.17) & 8.27 (0.06) & 8.80 (0.06)\\
\hspace{1em}$\theta(\textbf{s}_j)$ & 0.31 (0.02) & 0.18 (0.00) & 0.18 (0.00)\\
\hspace{1em}$\text{z}\left\{\textbf{s}_j; \delta(\textbf{s}_j)\right\} \theta\left(\textbf{s}_j\right)$ & 0.13 (0.00) & 0.55 (0.01) & 0.58 (0.01)\\
\hspace{1em}$\beta_1$ & 0.65 (0.00) & 0.50 (0.00) & 0.49 (0.00)\\
\hspace{1em}$\gamma_1$ & -- & 0.36 (0.01) & 0.52 (0.01)\\
\hspace{1em}$\eta_1$ & -- & -- & 0.14 (0.01)\\
\addlinespace[0.3em]
\multicolumn{4}{l}{\textbf{Misspecified}}\\
\hspace{1em}$\delta(\textbf{s}_j)$ & 0.94 (0.06) & 5.71 (0.06) & 5.84 (0.06)\\
\hspace{1em}$\theta(\textbf{s}_j)$ & 0.11 (0.01) & 0.14 (0.00) & 0.19 (0.00)\\
\hspace{1em}$\text{z}\left\{\textbf{s}_j; \delta(\textbf{s}_j)\right\} \theta\left(\textbf{s}_j\right)$ & 0.08 (0.00) & 0.31 (0.01) & 0.38 (0.01)\\
\hspace{1em}$\beta_1$ & 0.52 (0.00) & 0.49 (0.00) & 0.49 (0.00)\\
\hspace{1em}$\gamma_1$ & -- & 0.25 (0.00) & 0.32 (0.01)\\
\bottomrule
\end{tabular}
\end{table}
\clearpage

\begin{table}
\centering
\caption{Proportion of the $500$ simulated datasets favored by each competing method in the \textbf{Gaussian outcome} simulation, with respect to Watanabe-Akaike information criteria (WAIC), the Pareto-smoothed importance sampling leave-one-out (PSIS-LOO) cross validation, and the logarithm of the pseudo marginal likelihood (LPML), across all simulation settings. (Note, proportions may not sum to one due to rounding).}\label{table:sim_gauss_lowest_waic}
\begin{tabular}[t]{lccc}
\toprule
Simulation Setting & SVBR$\left[\delta, \theta\right]$ & SVBR$\left[\delta\left(\textbf{s}_j\right), \theta\right]$ & SVBR$\left[\delta\left(\textbf{s}_j\right), \theta\left(\textbf{s}_j\right)\right]$\\
\midrule
\textbf{No Effect}               &      &      &     \\
\hspace{1em} WAIC                & 0.38 & 0.43 & 0.19\\
\hspace{1em} LPML           &   0.37   &   0.45   &   0.18  \\
\hspace{1em} PSIS-LOO                &    0.37  &   0.45   &  0.18   \\
\textbf{Single Radius \& Effect}          &      &      &     \\
\hspace{1em} WAIC                & 0.71 & 0.17 & 0.12\\
\hspace{1em} LPML            &    0.72  &   0.17   &  0.11   \\
\hspace{1em} PSIS-LOO                &  0.72  &   0.17   &  0.12    \\
\textbf{Varying Radii, Single Effect}   &      &      &     \\
\hspace{1em} WAIC                & 0.00 & 0.73 & 0.27\\
\hspace{1em} LPML            &  0.00    &  0.73    &  0.27   \\
\hspace{1em} PSIS-LOO                & 0.00    &  0.72    &  0.28   \\
\textbf{Varying Radii \& Effect}          &      &      &     \\
\hspace{1em} WAIC                & 0.00 & 0.16 & 0.84\\
\hspace{1em} LPML            &    0.00  &  0.14    &  0.86   \\
\hspace{1em} PSIS-LOO                &   0.00   &  0.15    & 0.85    \\
\textbf{Misspecified} &      &      &     \\
\hspace{1em} WAIC                & 0.00 & 0.60 & 0.40\\
\hspace{1em} LPML            &   0.00   &   0.60   & 0.40    \\
\hspace{1em} PSIS-LOO                &    0.00  &  0.59    &  0.41   \\
\bottomrule
\end{tabular}
\end{table}
\clearpage

\section{Application study results}

\begin{table}[h!]
\centering
\renewcommand{\arraystretch}{1.3}
\caption{Posterior summaries from each competing method applied to the Madagascar (Toliara Province) antenatal care case study data using the counts exposure definition. Posterior medians and 95\% highest posterior density intervals are reported.  Covariate results from the main regression model are reported on the odds ratio scale. Elevation, Wealth, and GHFI (Global Human Footprint Index) were each standardized prior to analysis (zero mean and variance of one) with the original standard deviation (SD) given in parentheses for interpretation purposes.}\label{table:application_exp0_all_posteriors}
\resizebox{\textwidth}{!}{
\centering
\begin{tabular}[t]{lcccc}
\toprule
\multicolumn{1}{c}{ } & \multicolumn{4}{c}{Model Type} \\
\cmidrule(l{3pt}r{3pt}){2-5}
Parameter & SVBR$\left[5, \theta\right]$ & SVBR$\left[\delta, \theta\right]$ & SVBR$\left[\delta\left(\textbf{s}_j\right), \theta\right]$ & SVBR$\left[\delta\left(\textbf{s}_j\right), \theta\left(\textbf{s}_j\right)\right]$\\
\midrule
\multicolumn{1}{l}{\textbf{Main Regression Model}} & \multicolumn{4}{c}{} \\
Age (Years):                                                             &                   &                      &                              &                             \\
\hspace{1em}[20–35) vs.\ [12-20)                                          & 1.25 (0.91, 1.65) &  1.20  (0.88,  1.60) &   1.18   (0.84, 1.58)        &   1.17    (0.83, 1.57)\\
\hspace{1em}[35–50) vs.\ [12-20)                                          & 1.00 (0.64, 1.40) &  0.97  (0.65,  1.39) &   0.92   (0.58, 1.31)        &   0.92    (0.58, 1.33)\\
Education:                                                                &                   &                      &                              &                  \\
\hspace{1em}Primary vs.\ None                                             & 1.33 (1.05, 1.64) &  1.29  (1.02,  1.60) &   1.24   (0.96, 1.55)        &   1.22    (0.94, 1.53)\\
\hspace{1em}Secondary or Higher vs.\ None                                 & 2.79 (1.98, 3.84) &  2.61  (1.81,  3.52) &   2.28   (1.54, 3.17)        &   2.28    (1.50, 3.12)\\
Employed vs.\ Unemployed                                                  & 1.45 (1.08, 1.89) &  1.54  (1.15,  2.03) &   1.51   (1.10, 2.00)        &   1.52    (1.11, 2.03)\\
Married/Cohabiting vs.\                                                   & 1.25 (0.99, 1.56) &  1.24  (0.98,  1.53) &   1.26   (0.98, 1.58)        &   1.26    (0.98, 1.57)\\
Never/Widowed/Divorced/Separated                                          &                   &                      &                              &                  \\
First Birth vs.\ Not First Births                                             & 1.31 (0.94, 1.75) &  1.28  (0.91,  1.69) &   1.27   (0.88, 1.72)        &   1.28    (0.89, 1.74)\\
Religion:                                                                 &                   &                      &                              &                    \\
\hspace{1em}Other Religon vs.\ Christian                                  & 0.33 (0.14, 0.52) &  0.36  (0.19,  0.57) &   0.43   (0.21, 0.71)        &   0.46    (0.23, 0.76)\\
\hspace{1em}None vs.\ Christian                                           & 0.55 (0.44, 0.68) &  0.57  (0.45,  0.70) &   0.52   (0.40, 0.66)        &   0.54    (0.41, 0.68)\\
Urban Residence vs.\ Rural                                                & 0.59 (0.40, 0.81) &  0.48  (0.32,  0.66) &   0.27   (0.12, 0.44)        &   0.26    (0.10, 0.46)\\
Region:                                                                   &                   &                      &                              &                   \\
\hspace{1em}Anosy vs.\ Androy                                             & 0.76 (0.56, 0.98) &  0.81  (0.60,  1.04) &   0.59   (0.37, 0.86)        &   0.58    (0.38, 0.85)\\
\hspace{1em}Atsimo-Andrefana vs.\ Androy                                  & 0.87 (0.65, 1.14) &  0.80  (0.58,  1.04) &   0.73   (0.45, 1.08)        &   0.71    (0.46, 1.03)\\
\hspace{1em}Menabe vs.\ Androy                                            & 0.51 (0.38, 0.67) &  0.55  (0.41,  0.73) &   0.42   (0.24, 0.67)        &   0.44    (0.25, 0.73)\\
\multicolumn{1}{l}{\textbf{Buffer Radii Regression Model}} & \multicolumn{4}{c}{} \\
Radius (km) ($\delta\left(\textbf{s}_i\right)$)                           & 5.00 (Fixed)      & 12.33 (10.49, 15.01) & Varies across $\textbf{s}_j$ & Varies across $\textbf{s}_j$\\
Elevation (SD: 239.67)                                                                 & --                & --                   &  -0.10  (-0.39, 0.14)        &  -0.08   (-0.38, 0.21)\\
Wealth (SD: 0.31)                                                                    & --                & --                   &   0.07  (-0.13,   0.28)      &   0.13   (-0.12, 0.41)\\
GHFI (SD: 10.31)                                                                       & --                & --                   &  -0.23  (-0.41, -0.01)        &   0.20   (-0.10, 0.62)\\
$\tau_{\phi}$                                                & --                & --                   &   0.94   (0.81,   1.00)      &   0.91   (0.75, 1.00)\\
$\rho_{\phi}$                                               & --                & --                   &   8.28   (3.96,  14.00)      &   6.65   (2.54, 12.17)\\
Effective Range (km)                                         & --                & --                   & 290.14 (140.70, 505.17)      & 361.37 (154.74, 731.88)\\
\multicolumn{1}{l}{\textbf{Exposure Effect Regression Model}} & \multicolumn{4}{c}{} \\
Exposure Effect ($\theta\left(\textbf{s}_i\right)$)  & 1.16 (1.08, 1.25) &  1.17  (1.11,  1.24) &   1.86   (1.57, 2.18)        & Varies across $\textbf{s}_j$\\ 
Elevation (SD: 239.67)                                                                 & --                & --                   & --                           &  -0.04   (-0.18, 0.12)\\
Wealth (SD: 0.31)                                                                    & --                & --                   & --                           &  -0.01   (-0.12, 0.10)\\
GHFI (SD: 10.31)                                                                       & --                & --                   & --                           &  -0.15   (-0.28, -0.06)\\
\bottomrule
\end{tabular}
}
\end{table}

\clearpage

\begin{figure}[h]
    \centering
        \centering
        \includegraphics[width=1.00\textwidth]{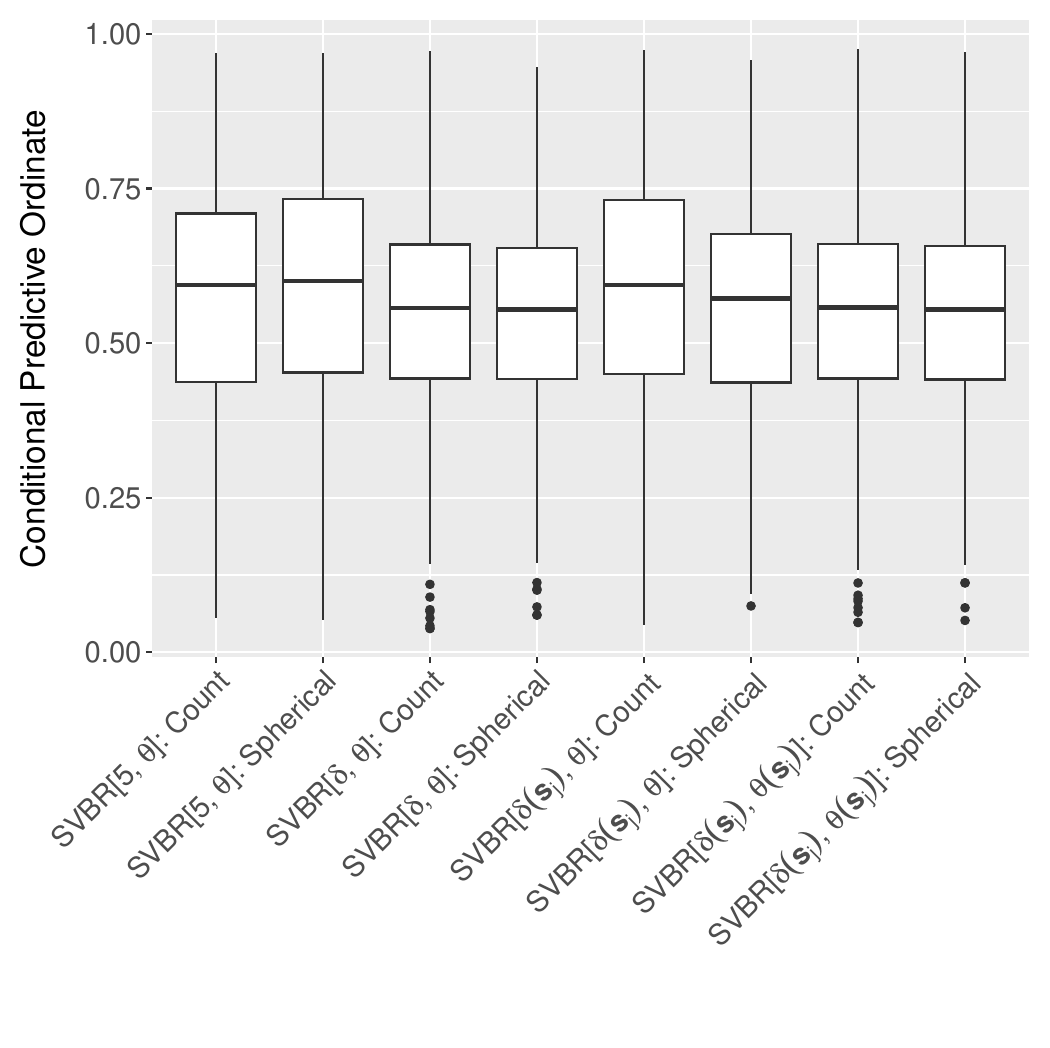}
    \caption{Conditional predictive ordinate box plots for each of the competing methods. All models are applied to the Madagascar (Toliara Province) antenatal care case study data using the counts exposure definition.}
\end{figure}
\clearpage

\begin{figure}[h]
    \centering
        \centering
        \includegraphics[width=0.49\textwidth]{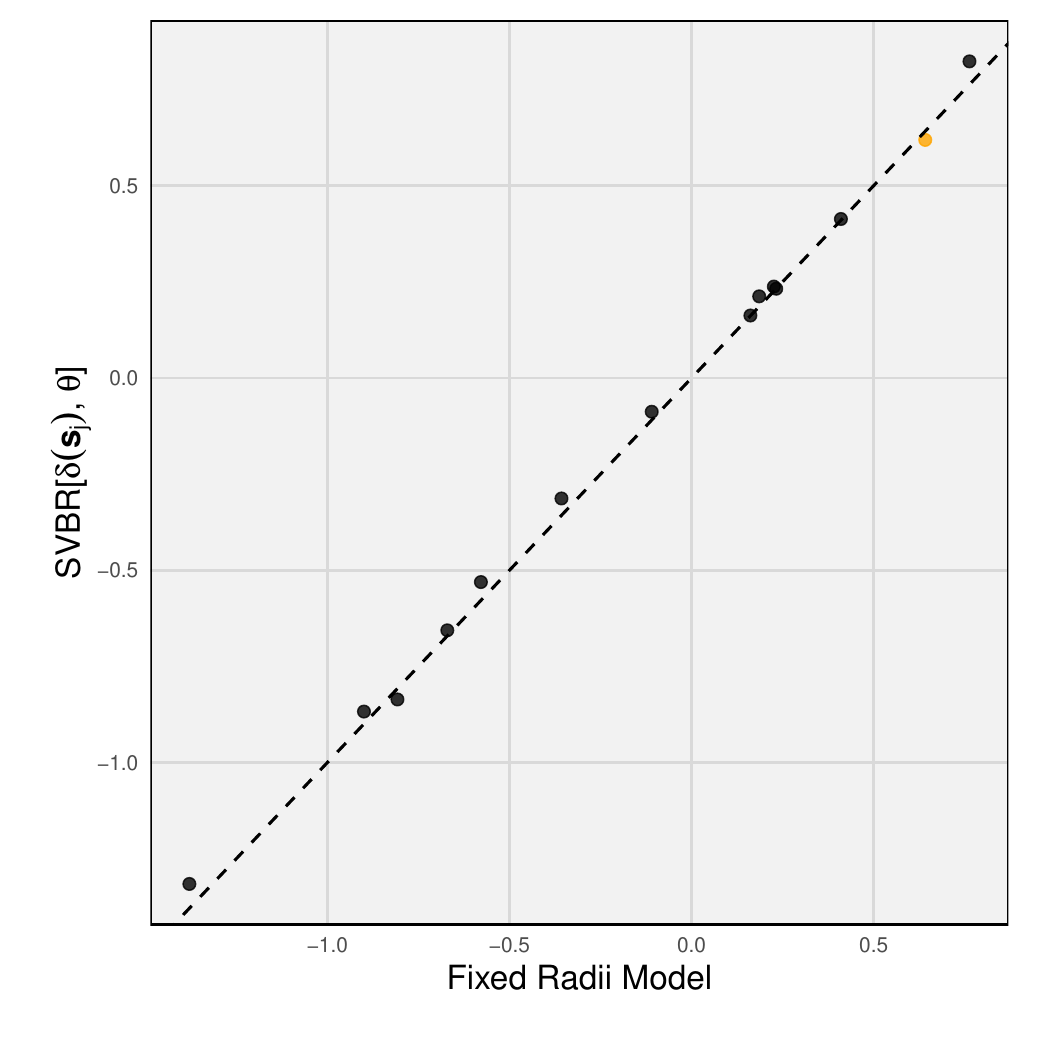}
        \includegraphics[width=0.49\textwidth]{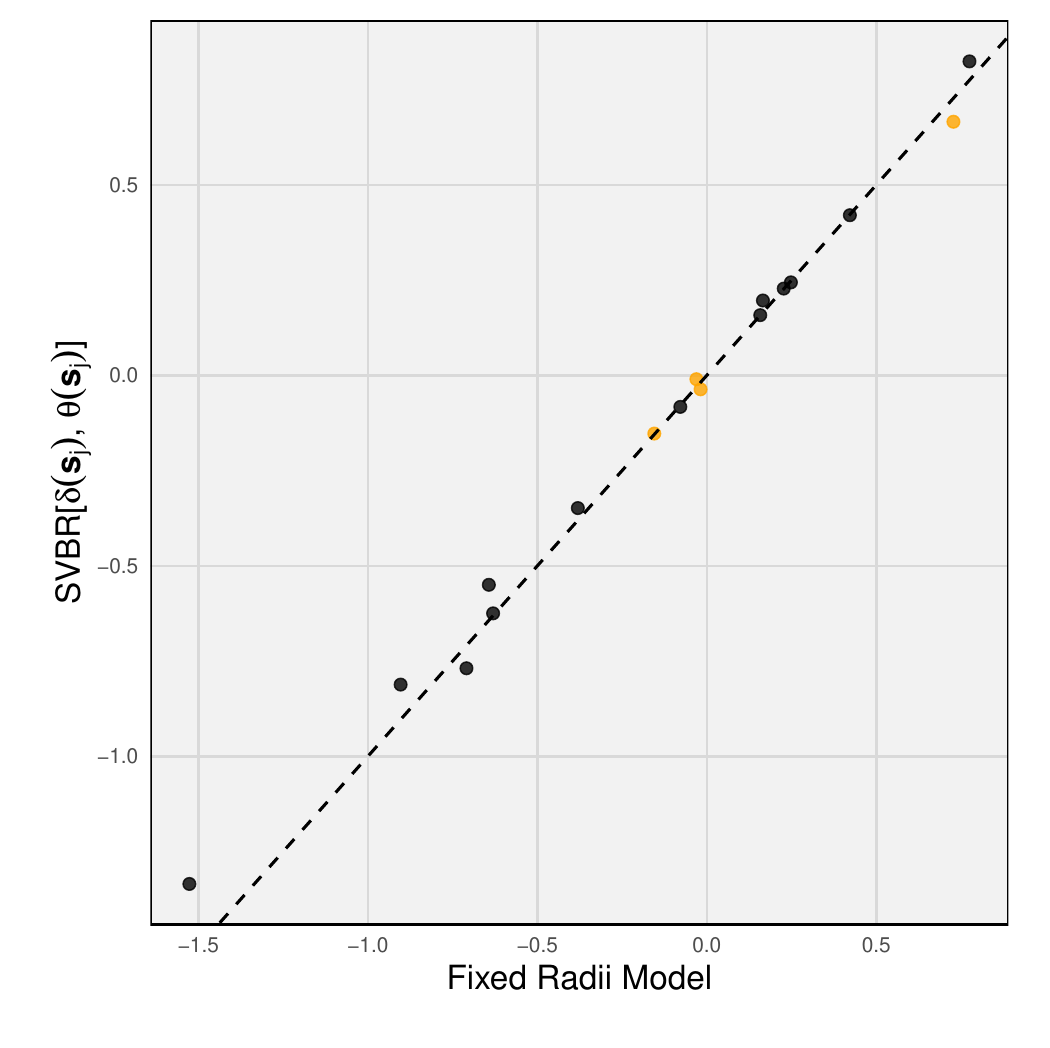}
    \caption{A comparison of $\boldsymbol{\eta}$ (orange) and $\boldsymbol{\beta}$ (black) point estimates (excluding the intercept in $\boldsymbol{\beta}$) from the joint Spatially-Varying Buffer Radii (SVBR) models and standard non-Bayesian logistic regression models that treat exposure as fixed/known using posterior median buffer radii estimates from the corresponding SVBR model. The dashed line represents the line of equality. All models are applied to the Madagascar (Toliara Province) antenatal care case study data using the counts exposure definition.}
\end{figure}
\clearpage

\begin{table}
\renewcommand{\arraystretch}{1.3}
\centering
\caption{Comparison of SVBR$\left[12.33, \theta\right]$ results to results from a model that directly accounts for the DHS survey design (i.e., \textbf{svyglm}). All results are reported on the odds ratio scale. Intervals for SVBR$\left[12.33, \theta\right]$ represent 95\% highest posterior density intervals and represent 95\% confidence intervals for \textbf{svyglm}.  Both models are applied to the Madagascar (Toliara Province) antenatal care case study data using the counts exposure definition.}\label{table:svy_frequentist_comparison}
\resizebox{\textwidth}{!}{
\centering
\begin{tabular}[t]{lcc}
\toprule
\multicolumn{1}{c}{ } & \multicolumn{2}{c}{Model Type} \\
\cmidrule(l{3pt}r{3pt}){2-3}
Parameter & SVBR$\left[12.33, \theta\right]$ & Fixed (\textbf{svyglm})\\ 
\midrule
\textbf{Radius }(km) ($\delta\left(\textbf{s}_i\right)$) & 12.33 (Fixed) & 12.33 (Fixed)\\   
\textbf{Exposure Effect }($\theta\left(\textbf{s}_i\right)$)  & 1.18 (1.12, 1.24) & 1.16 (1.09, 1.23)\\
Maternal Age:                    &   &    \\ 
\hspace{1em}[20–35) vs.\ [12-20) & 1.19 (0.87, 1.58) & 1.13 (0.80, 1.60) \\ 
\hspace{1em}[35–50) vs.\ [12-20) & 0.97 (0.64, 1.38) & 1.07 (0.74, 1.54) \\ 
Married/Cohabiting vs.\          & 1.24 (0.98, 1.54) & 1.23 (0.95, 1.59) \\ 
Never/Widowed/Divorced/Separated   &   &  \\ 
Education:                         &   &  \\ 
\hspace{1em}Primary vs.\ None   & 1.29 (1.01, 1.60) & 1.25 (0.94, 1.68) \\ 
\hspace{1em}Secondary or Higher vs.\ None   & 2.61 (1.83, 3.55) & 2.56 (1.63, 4.04) \\ 
Employed vs.\ Unemployed                 & 1.54 (1.14, 2.01) & 1.47 (0.98, 2.19) \\ 
First Birth vs.\ Later Births             & 1.27 (0.91, 1.70) & 1.27 (0.87, 1.86) \\ 
Religion:                                   &   &   \\ 
\hspace{1em}Other Religon vs.\ Christian     & 0.35 (0.19, 0.57) & 0.52 (0.20, 0.87) \\ 
\hspace{1em}None vs.\ Christian              & 0.56 (0.44, 0.70) & 0.56 (0.39, 0.81)  \\ 
Urban Residence vs.\ Rural               & 0.47 (0.32, 0.65) & 0.46 (0.30, 0.70) \\
Region:                                  &   &   \\ 
\hspace{1em}Anosy vs.\ Androy       & 0.80 (0.59, 1.04) & 0.72 (0.40, 1.28) \\ 
\hspace{1em}Atsimo-Andrefana vs.\ Androy   & 0.79 (0.58, 1.03) & 0.71 (0.41, 1.25) \\ 
\hspace{1em}Menabe vs.\ Androy            & 0.55 (0.40, 0.72) & 0.45 (0.26, 0.80) \\ 
\bottomrule
\end{tabular}
}
\end{table}

\end{document}